\documentclass[twocolumn,twocolappendix,dvipsnames]{aastex7}
\usepackage{amsmath}
\usepackage{multirow}
\usepackage{graphicx}
\usepackage{xspace}
\usepackage{subfigure}
\usepackage{changepage}
\usepackage{verbatim} 
\usepackage{textcomp}
\usepackage{gensymb}
\usepackage{float}
\usepackage{color}
\usepackage{silence}
\usepackage{hyperref}
\definecolor{ForestGreen}{rgb}{0.2,0.6,0.2}

\newcommand{\hMpc}{$h^{-1}$~Mpc\xspace}
\newcommand{\lth}{$\lambda_\mathrm{th}$}
\newcommand{\cmmnt}[1]{}

\begin{document}
\title{DeepVoid: A Deep Learning Void Detector}

\correspondingauthor{Michael Vogeley}
\email{vogeley@drexel.edu}

\author[0009-0004-2967-3847]{Sam Kumagai}
\affiliation{Department of Physics, Drexel University, Philadelphia, PA 19104}
\email{sk3993@drexel.edu}

\author[0000-0001-7416-9800]{Michael S. Vogeley}
\affiliation{Department of Physics, Drexel University, Philadelphia, PA 19104}
\email{vogeley@drexel.edu}

\author[0000-0003-0615-3975]{Miguel A. Aragon-Calvo}
\affiliation{Instituto de Astronomía, UNAM, Apdo. Postal 106, Ensenada 22800, B.C., México}
\email{maragon@astro.unam.mx}

\author[0000-0002-9540-546X]{Kelly A. Douglass}
\affiliation{Department of Physics \& Astronomy, University of Rochester, 500 Wilson Blvd., Rochester, NY  14627}
\email{kellyadouglass@rochester.edu}

\author[0000-0001-5537-4710]{Segev BenZvi}
\affiliation{Department of Physics \& Astronomy, University of Rochester, 500 Wilson Blvd., Rochester, NY  14627}
\email{segev.benzvi@rochester.edu}

\author[0000-0002-2618-5790]{Mark Neyrinck}
\affiliation{Department of Physics \& Astronomy, University of Denver, Denver, CO 80208}
\affiliation{Blue Marble Space Institute of Science, Seattle, WA 98104}
\email{mark.neyrinck@gmail.com}

%%% Abstract %%%
\begin{abstract}
    We present DeepVoid, an application of deep learning trained on a physical definition of cosmic voids to detect voids in density fields and galaxy distributions. By semantically segmenting the IllustrisTNG simulation volume using the tidal tensor, we train a deep convolutional neural network to classify local structure using a U-Net architecture for training and prediction. 
    The model achieves a void F1 score of 0.96 and a Matthews correlation coefficient over all structural classes of 0.81 for dark matter particles in IllustrisTNG with interparticle spacing of $\lambda=0.33$ \hMpc.
    We then apply the machine learning technique of curriculum learning to enable the model to classify structure in data with significantly larger intertracer separation.  At the highest tracer separation tested, $\lambda=10$ \hMpc, the model achieves a 
    void F1 score of 0.89
    and a Matthews correlation coefficient of 0.6 on IllustrisTNG subhalos.
\end{abstract}

%%% Sections %%%
% 0 introduction:
\section{Introduction}\label{sec:intro}
% Cosmic web, voids in galaxy surveys.
The large-scale structure of the Universe, as observed in galaxy redshift surveys, resembles a weblike structure \citep{Bond1996}. This structure is, by volume, chiefly comprised of cosmic voids, underdense regions that span tens of megaparsecs and are bounded by sheets and filaments of galaxies that flow into large galaxy clusters.  Depending on the definition, voids fill approximately $\sim$70\% of the volume of the Universe \citep{pan2012}. Voids were found in some of the earliest galaxy and cluster redshift surveys \citep[e.g.,][]{Gregory1978, Joveer1978} and were shown to be ubiquitous features of the large-scale structure by dense three-dimensional galaxy redshift surveys \citep{deLapperent1986, Geller1989, Shectman1996}.

Voids represent an interesting opportunity for cosmology and physics \citep{cai2025}. Voids grow from initial troughs in the primordial field of density fluctuations and continue to become more vacuous as the large-scale structure in the Universe evolves. They are ideal laboratories for studying modifications to general relativity \citep{perico2019} by observing, for example, the weak lensing signature of voids \citep{baker2018} or the gravitational redshift of cosmic microwave background (CMB) photons by large voids, known as the integrated Sachs-Wolfe (ISW) effect \citep{cai2014}.

Investigations of dark energy (DE) can also greatly benefit from studying voids. Because overdense structures such as galaxy clusters formed long before dark energy could strongly influence the evolution of large-scale structure, voids are the primary component of the cosmic web influenced by DE. Voids act as super-Hubble bubbles that expand faster than the Hubble flow \citep{Goldberg2004}. To leading order, the accelerated expansion of voids is the primary signature of DE; the overdense structures accelerate away from each other because the voids are expanding between them \citep{sheth2004}. 

Several specific cosmological tests using cosmic voids are of particular interest and motivate how our void finder is designed and implemented. The shapes of voids provide an avenue to test cosmological parameters through the Alcock-Paczyński (AP) effect \citep[e.g.,][]{alcock1979, Lavaux2012, hamaus2020}. The size spectrum of voids is sensitive to both the growth of structure and the DE equation of state \citep[e.g.,][]{Nadathur16, Versa19, Contarini2022, Contarini23}. Observation of the ISW effect from voids can further constrain cosmological models \citep[e.g.,][]{grannett2008,kovacs2018}.

One of the chief difficulties in finding and using voids for cosmology and astrophysics is the definition of a void. There is no universally agreed-upon definition, but studies generally rely on finding underdensities in the galaxy distribution, with methods mostly varying in the shape of the density estimation kernel. Popular void-finding methods for galaxy surveys utilize techniques that include 
the detection of voids as spherical regions or unions of multiple spheres \citep[e.g.,][]{Kauffmann91, ElAd97, Hoyle02, Banerjee2016, Zhao2016, Paz2023}, and watershed segmentation of the galaxy density field \citep[e.g.,][]{10.1111/j.1365-2966.2007.12125.x, Neyrinck08, Sutter15, Nadathur19b}.
Void catalogs constructed from identical galaxy samples can differ drastically based on how a void is defined in that study \citep{Colberg08}. 
These different definitions make it complicated to compare analyses based on different void catalogs.
Many of these methods also rely to varying degrees on fine-tuned parameters and different a priori notions of what constitutes a void and how they should appear. 

In this work, we demonstrate a proof of concept of an alternative approach: define voids in terms of the physics that creates and shapes them, and train a deep neural network to recognize the spatial signature of that physical process.
Here we describe DeepVoid, a void detector based on physical criteria to identify voids using convolutional neural networks (CNNs).
By employing such criteria, we can train CNNs to detect voids that are useful for cosmological tests of interest.
Previous investigations of using deep learning for morphological classification include \citet{aragoncalvo2019} and \citet{Inoue2022}. Unlike previous work, DeepVoid is targeted to the specific task of detecting voids in galaxy redshift surveys (although, as we will show, it can identify all the components of the cosmic web), a task which grows more challenging as these surveys probe deeper into the Universe.

The eventual goal of this approach is to fully leverage the cosmological information available in current and upcoming galaxy surveys and instruments, such as the Dark Energy Spectroscopic Instrument \citep[DESI;][]{DESI2016a.Science}, the Vera C. Rubin Legacy Survey of Space and Time \citep[LSST;][]{Abell09}, Euclid \citep{Mellier24}, the Nancy Grace Roman Space Telescope \citep{Schlieder24}, and the Subaru/Hyper Suprime Camera \citep[HSC;][]{Miyazaki18}. These surveys and others will provide maps of large-scale structure to unprecedented depths and with remarkable precision. They will probe ranges of cosmological redshift over which the evolution of voids is apparent and over which details of DE properties, possible modifications to general relativity, and neutrino properties might be probed \citep{Massara2015, Pisani15, perico2019, Avsajanishvili2023}.

Important challenges include that the galaxy catalogs from these new surveys may be quite sparse compared to more local galaxy surveys and that large-scale structure is probed by many different types of galaxy tracers with different biases \citep{Scaramella2022, Wang2022, desi_bgs, Zhou2023, Raichoor2023, Chaussidon2023}. 
The effect of ``tracer bias" on voids has been shown to affect the void size distribution 
\citep{Nadathur2015, Contarini2019}, impact the tracer profile vs. matter profile in voids \citep{Pollina2017}, and alter the velocity profile in voids, which is important for using void-galaxy correlations to study redshift-space distortions \citep[RSDs;][]{Massara2022}.
Current methods of estimating the void spectrum involve corrections for that bias applied after making a void catalog, based on analysis of mock galaxy catalogs \citep{Contarini2022}.
We will minimize the need for such postdetection void corrections by using mock catalogs to train variants of DeepVoid for specific galaxy target classes.
In this proof-of-concept paper, we demonstrate an approach to tackling the sparse sampling problem. In a follow-up paper, we will examine how the deep learning approach addresses the variety of galaxy tracers.

This study is timely, as machine learning and artificial intelligence methods have become commonplace in many fields of study and have been applied to nearly every domain in cosmology and astrophysics. Myriad examples of use cases include: segmenting the cosmic web into filaments and walls \citep{aragoncalvo2019}, building realizations of the cosmic web cheaply using generative networks \citep{rodriguez2018}, reconstructing initial conditions from galactic luminosities and positions \citep{modi2018}, studying the formation of dark matter (DM) halos \citep{luciesmith2019}, recovering the underlying DM distribution from galaxy positions \citep{zhang2019}, and extracting cosmological parameters from the matter distribution \citep{ravan2017,pan2020}. 

% roadmap of this paper
In this work, we present DeepVoid, a novel multiclass method of segmenting the morphology of structures in cosmological simulations using CNNs trained to match the classes assigned from properties of the gravitational potential. This paper is organized as follows. In Section~\ref{sec:simulations}, we describe the simulations from which we define our truth table. The physical definition of a void used to generate the truth table is discussed in Section ~\ref{sec:ttensor}. Aspects of deep learning and the U-Net architecture are detailed in Section~\ref{sec:deep_learning}. Model training and performance are presented in Sections~\ref{sec:training} and \ref{sec:anal}, respectively.
Training and performance of models that predict on sparse samples of halos are described in Section~\ref{sec:sparse}.
Conclusions and thoughts about future work are shared in Section~\ref{sec:disc}. Additional details on 
multiclass performance metrics, 
our use of batch normalization, and different loss functions can be found in the Appendices.

% 1 simulation details
\section{Simulation Data}\label{sec:simulations}
To define a truth table based on the underlying physics that shapes voids and the cosmic web at large, we need simulated datasets that are large enough to contain many instances of voids. CNNs need numerous examples to learn these patterns, so that if a model ``sees'' enough different cases during training, it can sufficiently generalize and therefore predict accurately when given novel data or instances. To confirm that our model is learning to identify structural morphology from the density field and not simply memorizing the layout of a particular simulation, we reserve a subset of the simulation volume to use solely for testing purposes, referred to herein as the validation set.

% simulation data: TNG and TNG300 specifications
The work presented here is based on The Next Generation Illustris Simulation (TNG) \citep{springel2018, pillepich2018, nelson2018, naiman2018, marinacci2018}. TNG is a gravo-magnetohydrodynamical simulation run using the moving-mesh code AREPO \citep{springel2010,vogelsberger2014}. The TNG suite is made up of three volumes of different sizes: 35, 75, and 205 \hMpc on a side. All TNG runs were computed with the Planck 2015 cosmology: $\Omega_{\Lambda}=0.6911$, $\Omega_{m}=0.3089$, $\Omega_{b}=0.0486$, $\sigma_8 = 0.8159$, $n_s = 0.9667$, and $h = 0.6774$ \citep{planck2016}. 

For this work, we use the largest volume, TNG300, which is 205 \hMpc on each side. 
TNG300 has three different realizations, with TNG300-1 having $2500^3$ DM particles, TNG300-2 having $1250^3$ DM particles, and TNG300-3 having $625^3$ DM particles. We use the $z=0$ snapshot of TNG300-3, but there are 127 snapshots going back to $z=127$ (with the same redshifts as the original Illustris simulation). The proof-of-concept models described in this paper are trained on either the spatial distribution of DM particle mass density or that of the DM subhalos. Since we are only interested in the distribution of matter on the largest scales, we determined that the DM-only versions of the TNG300 volumes were sufficient for our purposes. TNG300-3-DARK has a mass resolution of $M_{\rm DM}=3\times 10^9 h^{-1}~M_\odot$. In the future, as we continue to enhance our void detectors, we plan to incorporate additional information from simulations, such as galaxy luminosity and color, to improve predictive performance. 

% getting full DM density and halo catalogs
To train on data sets that emulate the sparseness of galaxy redshift surveys, we use DM halo samples with number densities comparable to the Sloan Digital Sky Survey \citep[SDSS;][]{SDSS7} main galaxy sample or to the DESI Bright Galaxy Survey \citep[e.g.,][]{desi_bgs}. We use the subhalo catalog that was generated using the SUBFIND algorithm \citep{springel2001}, a computational tool to identify and analyze substructures in DM halos. These subhalos are smaller gravitationally-bound clumps of matter within each halo.

In this work, ``interparticle'' spacing refers to the mean spatial separation of DM particles of the respective simulation. If we refer to that of the subhalos, we use the term ``intertracer'' spacing or separation. In both cases, we define a characteristic interparticle/intertracer spacing $\lambda \equiv n^{-1/3}$, where $n = (N/V)$ is the tracer density. To form samples of halos as tracers, we sort them by virial mass, then select $N_{\rm tracers}$ of decreasing mass such that the intertracer separation $\lambda$ is what we desire. Although these are not perfect analogs for the galaxies we observe in surveys, they serve as a rough proxy for a volume-limited sample that traces the large-scale structure.

% 2 tidal tensor definition
\section{Tidal Tensor Definition}\label{sec:ttensor}

\begin{figure*}
    \centering
    \includegraphics[width=\textwidth]%{figures/new/TNG_d_p_m_comp_slc_255_zoom_100-FIX.pdf}
    {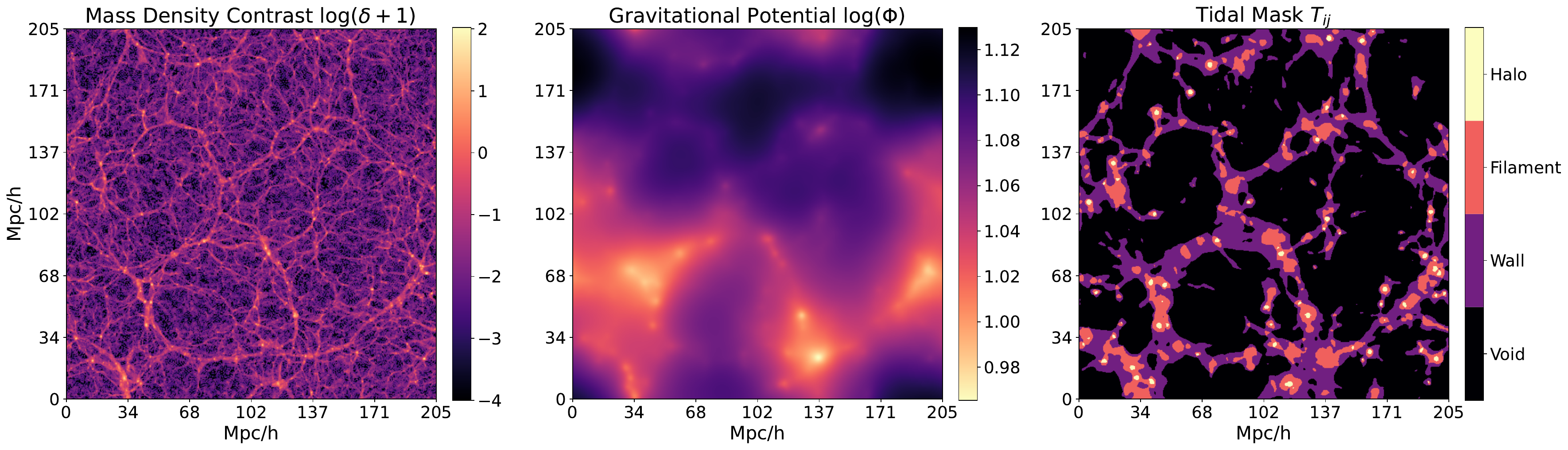}
    \caption{\emph{Left:} TNG300-3-Dark mass density contrast defined as $\delta(\Vec{x})=\rho(\Vec{x})/{\Bar{\rho}}-1$, where $\Bar{\rho}$ is the average matter density in the volume. \emph{Center:} corresponding gravitational potential $\Phi$. Note that both $\delta$ and $\Phi$ have been scaled and shifted such that their minimum value is zero and their standard deviation is one, and their logarithm plotted for ease of interpretability. \emph{Right:} multilabel tidal tensor classified mask. This mask is our ``truth table'' for training. These plots are one slice thick, with $N_{\rm mesh}=512$, so they are 0.4 \hMpc thick.}
    \label{fig:d_p_mask}
\end{figure*}

The best definition of cosmic voids and the optimal method for finding them may depend on the astrophysical or cosmological question of interest. For example, if the subject of interest is the difference in evolution between void and nonvoid galaxies, how well the definition delineates between void and wall regions is of paramount importance \citep{Douglass2023,Veyrat2023,Zadouni2024}. On the other hand, if one wants to investigate cosmological effects such as the impact of voids on the CMB due to lensing or through the ISW effect, then obtaining accurate locations of void centers with respect to the potential is more important \citep{Cai2017}.

For training our proof-of-concept deep learning void detector, we employ a definition of voids that is based on simple physics: the gravitational potential that drives the formation of large-scale structure.
Thus, we define structural morphology with the so-called ``T-web'' method, which was first used to segment large-scale structures by \citet{Hahn2007} \citep[see also][]{forero2009}. This method is based on the tidal tensor, the Hessian of the gravitational potential:
\begin{equation}
    T_{ij}(\Vec{x}) = \frac{\partial^2 \Phi(\Vec{x})}{\partial x_i \partial x_j}.
\end{equation}
\citet{Hahn2007} then found the eigenvalues of the tidal tensor and counted the number of positive eigenvalues to classify the environment of each voxel. 
Motivated by Zel'dovich's theory \citep{Zeldovich1970},
they defined voids as regions where $T_{ij}$ has no positive eigenvalues (unstable orbits), walls as regions with one positive eigenvalue (one-dimensional stable manifold), filaments having two positive eigenvalues (two-dimensional stable manifold), and finally halos as regions with all three positive eigenvalues (attractive fixed points).
The tidal tensor is a good definition for our purposes because it captures the local gravitational dynamics in a region in an intuitive and relatively simple way. Furthermore, the T-web method uses only two free parameters: 
% both of which are defined below: 
the smoothing scale, $\sigma$, and the eigenvalue threshold, $\lambda_\mathrm{th}$, used for distinguishing between the classes.

For the purpose of identifying voids, the tidal tensor definition is particularly useful because it describes the spatial pattern of void regions around peaks of the potential. 
By using the tidal tensor as the truth table upon which the network is trained in the DeepVoid model described below, voids are identified not simply as regions of low density, but also based on the spatial pattern of matter that is predicted to arise from gravity.

The tidal tensor definition of structural segmentation has been widely applied to simulations of large-scale structure, in which one has access to the exact matter density field, and so one can compute the gravitational potential and its derivatives using fast Fourier methods.
This definition is difficult to apply to observational data because the gravitational potential is not directly observable and depends on long-wavelength modes of the matter density field. 
Estimation of the potential and its derivatives from galaxy survey data is further limited by the survey boundaries, sparseness of tracers of the matter density, biasing of tracers, and RSDs. 
\cite{Wang2012} estimate the tidal tensor within the SDSS main galaxy sample region and find acceptable results only very far from the survey boundaries. \cite{Eardley2015} examine how estimation of the tidal tensor is influenced by the survey geometry, galaxy sampling, and redshift space distortions in application to the GAMA survey.

Related methods of classifying structural morphology include the use of the Hessian of the density field \citep{Aragon-Calvo2007}, the use of the Hessian of the velocity shear tensor \citep[the ``V-web method'';][]{Hoffman2012}, and 
ORIGAMI \citep{FalckEtal2012}, which counts stream crossings that have already occurred. 
\citet{Libeskind2018} review and compare methods of identifying structural elements of the cosmic web. Among their findings is that agreement between methods is highest for void regions (as opposed to denser structures, where some methods substantially disagree). The deep learning architecture we use, as described in Section~\ref{sec:deep_learning}, could be applied to other methods of constructing the truth table for training, so future versions of DeepVoid could be trained to leverage different definitions of the morphology for different cosmological purposes.

% Calculating potential, tidal tensor eigenvalues
To perform a classification using the tidal tensor, we must calculate the gravitational potential $\Phi$. First, we compute the matter density contrast $\delta(\Vec{r})$ by using the cloud-in-cell (CIC) interpolation scheme \citep{hockney1988} to allocate each DM particle's mass appropriately in a grid with $N_{\rm mesh}^3$ voxels, where $N_{\rm mesh}$ is the number of voxels on a side. Then, by using fast Fourier transforms (FFTs) and a Green's function to solve Poisson's equation,
\begin{equation}
    \hat{\Phi}(\Vec{k}) = - 4 \pi G \frac{\hat{\rho}(\Vec{k})}{|\Vec{k}|^2}, 
\end{equation}
we obtain the gravitational potential $\hat{\Phi}(\Vec{k})$ in Fourier space, where $\Vec{k}$ is the wavevector and hats denote values in Fourier space. We can then inverse-FFT this to obtain the potential in the spatial domain, $\Phi(\Vec{x})$. In Appendix~\ref{sec:tt_derive} we provide more mathematical details of the Fourier method for computing the potential.

Before computing the tidal tensor, some smoothing is necessary to remove possible high-frequency contributions to the derivatives of the potential (Forero-Romero et al. 2009).
Smoothing is also desirable because we want to segment large-scale structure based on relative accelerations on scales larger than that of individual halos (galaxies). 
To accomplish both goals, we smooth the potential on an effective smoothing scale $\sigma_{\rm eff}=$ 1\hMpc,
This smoothing includes both the intrinsic smoothing caused by binning particles onto the grid, $\sigma_{\rm grid}=L_{\rm box}/N_{\rm mesh}$ (for the TNG300 simulation, $N_{\rm mesh}=512$, thus each voxel is 0.4 \hMpc~ on a side) and an explicit Gaussian smoothing with scale $\sigma_G$ that is set to achieve the desired effective scale, $\sigma_{\rm eff}^2=\sigma_G^2 + \sigma_{\rm grid}^2$.
Use of a smaller smoothing scale would risk employing noisy estimates of the derivatives and focus attention on local structures rather than the large-scale features. A significantly larger smoothing scale would blur the boundaries between the morphological structures.

% figure: compare density, velocity, potential, TT segmentation by eigenvalues
\begin{figure*}
    \centering
    \includegraphics[width=0.95\textwidth]{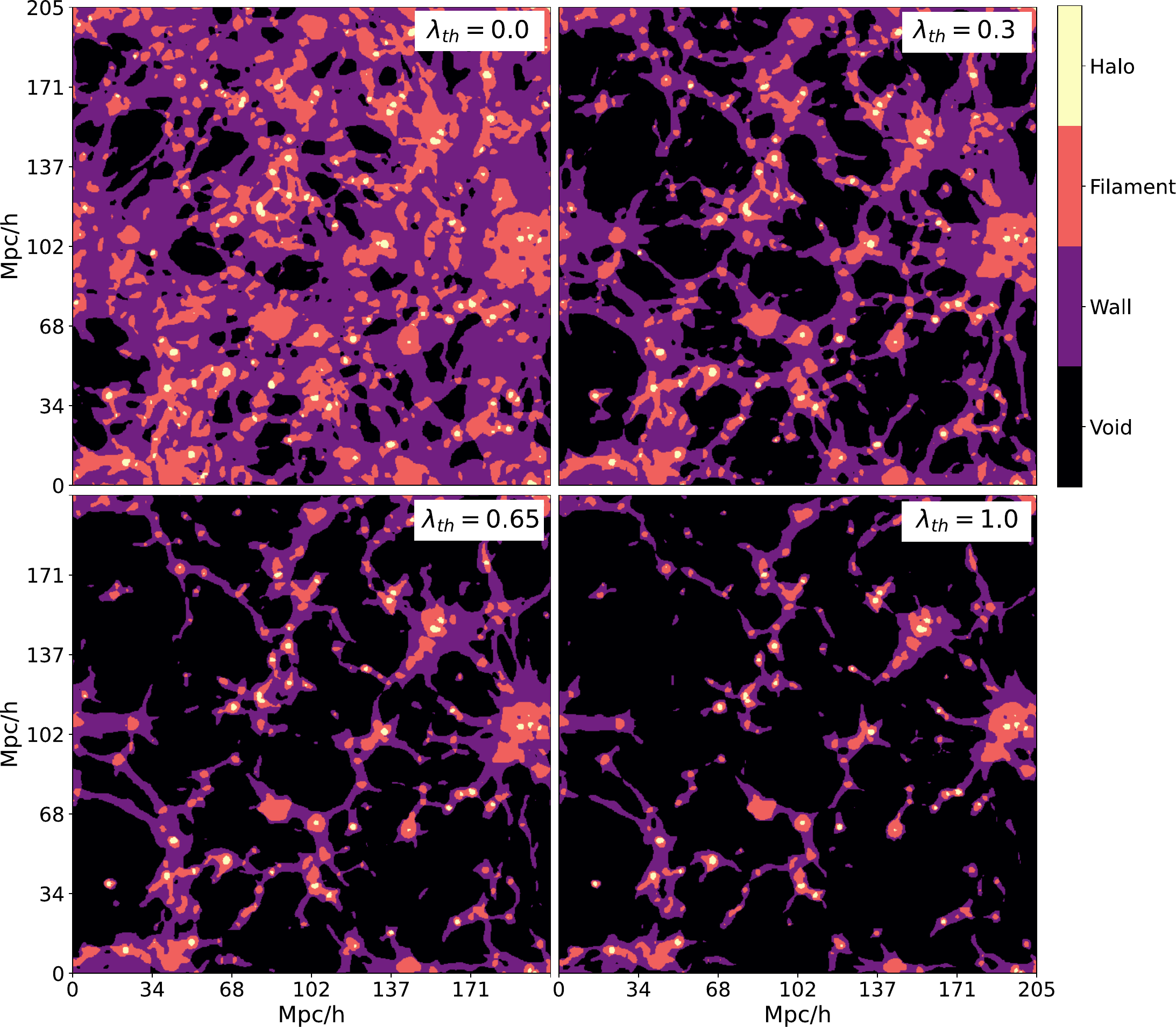}
    \caption{A comparison of different eigenvalue thresholds used to create a tidal-tensor-based map of structural morphology in TNG300-3-Dark. \emph{Top left:} $\lambda_\mathrm{th}=0$; \emph{Top right:} $\lambda_\mathrm{th}=0.3$; \emph{Bottom left:} $\lambda_\mathrm{th}=0.65$ (our default); \emph{Bottom right:} $\lambda_\mathrm{th}=1$. The morphology of the cosmic web as defined by \lth$=0$ consists of tiny isolated voids surrounded by huge amounts of wall voxels. We see a more realistic depiction of the Universe's large-scale structure at slightly positive values of \lth.}
    \label{fig:mask_comp}
\end{figure*}

Finally, we compute the Hessian of the potential to obtain the tidal tensor, $T_{ij}(\Vec{x})$. For each voxel, $T_{ij}$ contains six independent values due to symmetry. We then solve for the three eigenvalues for each voxel, and determine the class based on how many eigenvalues are greater than $\lambda_\mathrm{th}$. The result of performing this procedure on the TNG300 mass density field is shown in Figure~\ref{fig:d_p_mask}. 

One notable change to the method of \citet{Hahn2007} is our use of a nonzero eigenvalue threshold $\lambda_\mathrm{th}$, as recommended in previous studies using similar techniques \citep{forero2009,alonso2015}.
% choosing a threshold value for eigenvalues
If we apply the exact procedure of \citet{Hahn2007} to the TNG300 sample, we do not obtain a classification of the large-scale structure that matches what we observe in galaxy surveys or expect in simulations. If, as they did, we choose $\lambda_\mathrm{th}=0$ to differentiate between compressing and lengthening tidal forces, we find that only 16\% of the total TNG300 volume is classified as void. 

Choosing \lth$=0$ means that if an eigenvalue is even infinitesimally positive, the classification assumes that the locality around that grid point collapses along that eigenvector. This collapse happens over a dynamical timescale. Thus, for an infinitesimally positive value of the eigenvalue, collapse may either never occur (if the value is positive due to noise or transient behavior) or not until the remote future (possibly larger than the Hubble time). As \citet{forero2009} note, the dimensionality of $T_{ij}$ is [time]$^{-2}$, so its eigenvalues can be correlated with the freefall collapse time: 
\begin{equation}
    \tau_{ff} = \sqrt{\frac{3\pi}{32 G \rho}}\ .
\end{equation}
If we set the threshold such that the corresponding spherical collapse time is the same as the age of the Universe, we obtain this equation from Appendix A of \citet{forero2009}: 
\begin{equation} \label{eq:4}
    \lambda_\mathrm{th}=\frac{1}{3\beta(\lambda_i)}\left[ \frac{\pi^2}{4} \frac{1}{\Omega_m} (\tau_0 H_0)^{-2} -1 \right]\ ,
\end{equation}
where $\beta(\lambda_i)$ is a corrective factor that accounts for deviations from perfect isotropy. If we use TNG's Planck 2015 cosmological parameters and assume isotropy, $\beta(\lambda_i)=1$, then this order-of-magnitude calculation suggests a threshold for voids of $\lambda_\mathrm{th}=2.58$. 
\citet{alonso2015} argue that, because the density contrast is the sum of the eigenvalues, and nonlinear collapse occurs around a density contrast of unity, this eigenvalue threshold should be of order unity.
As other authors who have used the tidal tensor method discuss \citep{forero2009, alonso2015, Bonnaire2022}, the choice of value for this parameter depends on the properties of the cosmic web that are of interest. For example, if we were only concerned with finding the most underdense core regions of voids, then a threshold of zero would be useful.

Figure~\ref{fig:mask_comp} shows the effect of changing the eigenvalue threshold in this scheme. All training and testing wer done using tidal-tensor-based masks with $\lambda_\mathrm{th}=0.65$. We choose this value because it segments space into proportions of void, wall, filament, and halo morphologies that roughly match other methods. Given this threshold, TNG300 is composed of approximately 65\% void, 26\% wall, 9\% filament, and 0.5\% halo voxels.

% 3 deep learning architecture
\section{Application of Deep Learning}\label{sec:deep_learning}

\subsection{Why Use Deep Learning?}

Here, we describe the deep learning method that we adopt for translating the tidal tensor classification of large-scale structure into a method that can be applied to sparse galaxy survey data. As described above in Section~\ref{sec:ttensor}, if we were given the exact mass distribution in the Universe, as is true for cosmological simulations, then there would be no need for deep learning; one would simply compute the potential and its derivatives to form the tidal tensor. In the first application described below (Section~\ref{subsec:basemodels}), we build a deep learning model that is trained on the full DM density field and demonstrate that it accurately recovers the structural classes of the truth table that was also computed from that full DM distribution. In that specific test case, DeepVoid could be described as an ``emulator" of the structural class definitions that were computed from the tidal tensor. The need for deep learning arises when we move to training models appropriate for analysis of sparse galaxy data (Section~\ref{sec:sparse}). In this significantly more challenging application, the model must learn to recognize voids, filaments, walls, and halos from their spatial patterns imprinted on the distribution of these sparse and biased tracers.

\subsection{Convolutional Neural Networks}

Our choice of CNNs for void-finding is driven by their success in segmenting volume-filling structures in other three-dimensional data, particularly in the analysis of biomedical imaging \citep[e.g.,][]{NIYAS2022}.

CNNs employ convolutional layers that apply trainable filters to the input data, enabling the identification of spatially localized features. Additionally, through successive layers and pooling operations, CNNs can capture complex and nonlinear features. 

% Basics of CNN 
A CNN is built out of many layers of ``neurons,'' each of which takes in an input signal, multiplies it by an array of weights, adds a bias term, and that output is fed into an activation function. The activation function (described below) then produces the value for one pixel or voxel in the feature image. This operation is applied to all pixels/voxels in the input array by convolving a kernel of a chosen size with some stride. All networks we train use a $(3,3,3)$ kernel size in all convolutional layers. As is typical, we employ zero-padding on each convolutional block in our networks so that the stride and kernel size of the convolution span the volume, ensuring that the input and output feature images have the same dimensions. A deep CNN then feeds these feature images into the next layer as input. Multiple layers allow deep networks to encode different relevant aspects of the input data in nonlinear ways.

% activation fxn & MaxPooling3D
The activation function we employ is the rectified linear unit (ReLU) function: $\phi(x) = x$ if $x \geq 0$ and 0 if $x < 0$ \citep{glorot2011}. ReLUs allow for extremely efficient computation, as there are no arithmetic operations necessary. Using ReLU after all convolutions (except the last) enables the network to cheaply incorporate nonlinearity. MaxPooling layers in the CNN add nonlinear behavior as well, since they only pass down the maximum value of $\phi(x)$ from a compact set of pixels. We use MaxPooling3D layers with a kernel size of (2,2,2), a stride of 2, and ``same'' padding to ensure that the output maintains the same shape as the input. However, since the stride is equal to the kernel size, the spatial dimensions are effectively reduced by a factor of 2 after each pooling operation. At the same time, the number of feature channels doubles, allowing the network to learn progressively richer representations at increasingly coarser spatial scales.

Overfitting is an inherent challenge in training deep neural networks. Given a finite dataset, almost any sufficiently complex model will eventually learn not only meaningful patterns but also noise and random fluctuations specific to the training data, reducing its ability to generalize to new inputs. This issue becomes apparent when the training loss continues to decrease while the validation loss plateaus or increases, as seen in Figure~\ref{fig:epochs_vs_loss}. To mitigate overfitting, we employ several strategies. 
First, we augment our dataset in the form of $90\degree$ rotations, where each subcube is rotated three times, effectively increasing the  dataset size by a factor of 4. Additionally, we split the dataset into training and validation sets, ensuring that we are evaluating model performance on data that it has not seen before. 
We also experiment with different learning rates to balance convergence speed and model stability. We found that a learning rate of 0.0003 allowed the models to converge efficiently while still enabling effective learning. We employ a learning rate scheduler that quarters the rate if the validation loss has not improved in 15 epochs, allowing the models to refine their parameters more gradually, improving generalization. 
Finally, we save the model weights corresponding to the lowest validation loss during training, which ensures that we retain the best-performing model rather than one that may have overfitted in later epochs.

% Figure: U-net: number of levels, kernel size, stride, batch normalization, activation function, pooling
\begin{figure*}[t]
    \centering
    \includegraphics[width=\textwidth]{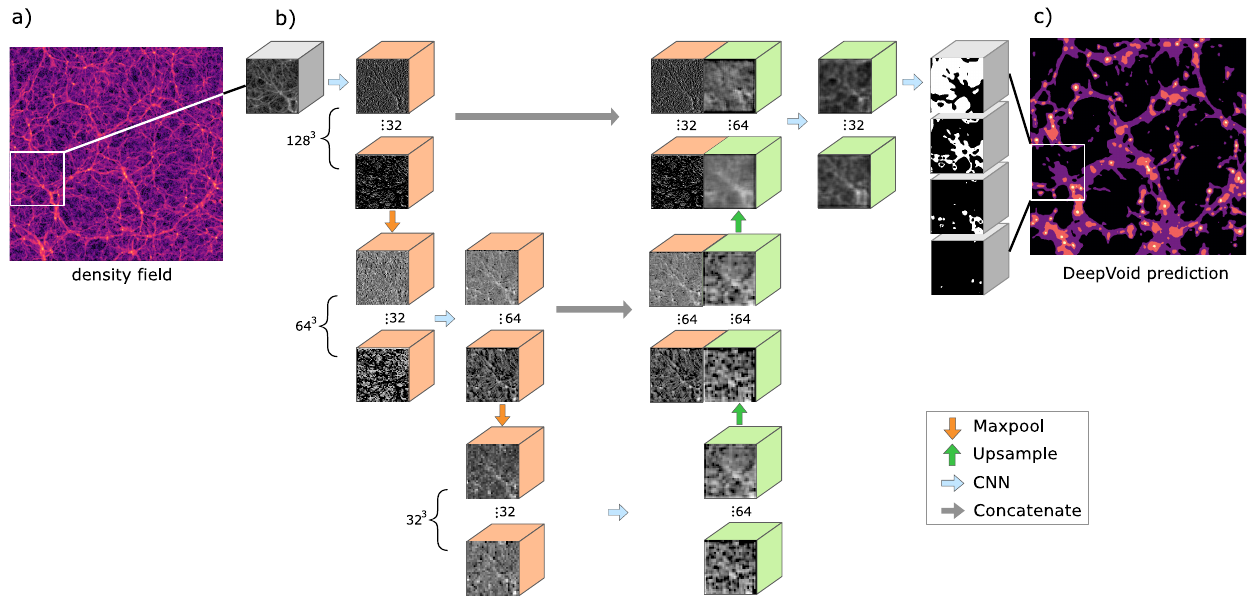}
    \caption{Schematic of a U-Net architecture with three levels of resolution (`depth'). In (a), chunks of a density field are fed through the successive layers in (b), with cubes representing 3D feature maps colored orange on the encoding side, and green on the decoding side. Numbers next to the curly brackets indicate the size of the activation images in a given layer, while numbers next to the vertical ellipses indicate the number of filters in each layer. Colored arrows indicate operations performed between different layers. Horizontal gray arrows represent the merging of feature maps used to transfer small-scale spatial information from the encoding layers to the decoding layers. Without these concatenations, the U-Net architecture is identical to that of an autoencoder. The predicted structural segmentation is shown in (c).
    }
    \label{fig:UNET}
\end{figure*}

% Details of U-Net architecture
\subsection{U-Net Architecture}\label{subsec:unet}

The architecture we use in this work is a U-Net, a type of CNN based on the model of \citet{Ronneberger2015}, originally designed to perform segmentation of neuronal structures imaged using electron microscopy. 
Similar to an autoencoder, the U-Net learns an internal compressed representation of the input as the information cascades down the encoding path. The data is then reconstructed by the network as the spatial resolution increases on its way up the decoding path. However, autoencoders suffer from a lack of locality. The U-net architecture addresses this by concatenating small-scale features from the encoding path to the decoding path. This allows the network to localize better while parsing data on multiple scales and has led to the U-net becoming one of the standard architectures for image processing networks. 
The U-Net architecture is named after the distinctive U-shaped network structure consisting of an encoding and decoding side. The encoder pathway, consisting of convolutional and pooling layers, progressively extracts higher-level features, enabling the network to learn a rich hierarchical representation of the input image. The decoder pathway, through upsampling and convolutional layers, recovers spatial details and combines them with corresponding features from the encoder pathway using concatenations. This facilitates the accurate localization of objects or regions in the segmentation output. 
For the problem of segmenting large-scale structure in the galaxy distribution, this neural network architecture allows us to train models that recognize the spatial patterns of voids, walls, filaments, and halos that gravity imprints on the galaxy distribution.
Figure~\ref{fig:UNET} illustrates the U-net architecture (see \cite{He19,Berger19, Jamieson23} for other applications of CNNs and UNet to 3D datasets in astronomy). 

Throughout the rest of this paper, we will refer to the number of encoding/decoding layers as the `depth' of the U-net, and the number of filters in the first convolutional block as the number of `filters' for that particular DeepVoid model (see Figure \ref{fig:UNET} for reference). 

As information propagates through the U-net, undergoing convolutions, poolings, upsamplings, and concatenations, it finally converges at the final convolutional layer. This layer produces a feature tensor with $N_{\rm classes}$ as the last dimension. This tensor is then fed to the last activation function. Since we are performing a multilabel classification, we set the last activation layer to the Softmax function: 
\begin{equation}
    \sigma(\Vec{z})_i = \frac{e^{z_i}}{\sum_{j=1}^K e^{z_j}} ,
\end{equation}
where $\Vec{z}$ is the input vector with components $(z_0,\ldots z_K)$ and $K$ is the number of classes \citep{bridlesoftmax}. The term in the denominator ensures that the probabilities of different classes for a pixel, or in our case, a voxel, always add up to one. This ensures that the output of the Softmax function resembles a probability distribution over $K$ different possible outcomes. Strictly speaking, the Softmax outputs are not probabilities, but rather are more like normalized confidence scores \citep{guo2017}. Finally, in order to predict a single class for each voxel, the Argmax function is used, which simply assigns the class with the greatest confidence score to each respective voxel.  

\subsection{Choosing a loss function}\label{subsec:losses}
% Details of training: loss function:
Once a model makes its predictions, a loss function is necessary to determine how well the predictions match the ground truth. In order to train a network, we need a differentiable loss function to minimize through backpropagation. The goal of training is to optimize the layer weights and biases such that it results in the lowest possible loss. The `loss landscape' is the high-dimensional space of those parameters, which is typically conceptualized as a surface. Ideally, the training results in the model finding a global minimum, i.e., as well as the model can perform given its hyperparameters. However, in practice, the optimizer is really trying to find the best possible local minimum in that landscape.

We choose a standard loss for multiclass semantic segmentation, sparse categorical cross entropy \citep[SCCE;][]{mao_cce}. SCCE, which is the same as the binary cross entropy loss function across multiple classes, can be represented by the following formula:
\begin{equation}
    L_{\rm CCE}(y,\hat{y}) = - \sum_{i}^N y_i \log(\hat{y_i}) , 
\end{equation}
where $y$ and $\hat{y}$ are vectors denoting the actual and predicted labels, respectively. The `sparse' part of SCCE indicates that the labels are encoded as integers instead of one-hot vectors, which greatly reduces the memory overhead during training. 

We experiment with two other loss functions as well, as described in more detail in Appendix~\ref{sec:loss_fxns}. Some other studies have found good results combining the Dice or F1 score with SCCE. When class imbalance is a priority, the focal loss, a modification of CCE, has also been successful. Some combination Dice/SCCE base models slightly outperform the SCCE base models listed in Table~\ref{table:base_models}. However, none of the combination Dice/SCCE or focal models did better than the SCCE models at larger intertracer spacings. Further tuning of the specific weightings of Dice/SCCE and the focal loss parameters could likely improve on these results, but that is beyond the scope of this proof-of-concept paper.

% 4 training and validation
\section{Training and Validation}\label{sec:training}
Training in this work was performed on two high-performance computing clusters: Picotte at the Drexel University Research Computing Facility and Nautilus at the National Research Platform. Each training run on Picotte used two NVIDIA Tesla V100 GPUs, each with 32 GB of memory, while predictions were performed on one Tesla V100 GPU. Training on Nautilus used various configurations of GPUs. We construct our 3D U-nets using the Keras library \citep{keras} that runs on top of the TensorFlow library \citep{tensorflow} in Python. Additionally, we use the Adam \citep[adaptive moment estimation,][]{kingma2014} stochastic gradient descent algorithm with an initial learning rate of 0.0003 and parameters $\beta_1=0.9$, $\beta_2=0.999$, and $\epsilon=10^{-7}$.

\subsection{Construction of training data}
%%% Construction of training/validation data sets
We construct training data by first splitting the density and mask fields into smaller subcubes. TNG's $512^3$ grid is broken into 343 $128^3$ subcubes. For data augmentation purposes and to help avoid overfitting, each subcube is rotated 3 times by 90\degree\, to take advantage of the lack of rotational invariance in CNNs. Therefore, the TNG dataset consists of 1372 subcubes. The resulting subcubes are then split into training and validation sets in an 80\%/20\%  split. The subcubes are also shuffled during training to avoid feeding the network batches of consecutive rotated subcubes. 

Note that all metrics reported in this work are computed using the validation dataset. 
We use the same validation data set to compute performance metrics when choosing hyperparameters, when computing validation loss during training, and for computing final performance metrics.
The fact that we do not use a separate testing dataset to evaluate final performance metrics is of possible concern if the examples in the validation dataset are not representative of the space of possible structures. That would also be true for a small separate testing dataset. Thus, we use a single validation/testing dataset to include a large set of different examples from the TNG300 simulation box. If we are unlucky and the examples in the validation data are quite dissimilar from the training data (assuming here that the training dataset is large enough to be representative), then this could bias both the ``stopping" point of training and the performance metrics: validation loss would exceed training loss too soon, causing an early stop to training, and result in poor performance metrics computed on the validation data, both because the training was stopped too soon and because the validation data are dissimilar to the training data.
Splitting TNG300 into smaller, separate training/validation/testing datasets would increase the probability of such a result.
Another possible concern is that using the same validation data to set hyperparameters admits the possibility that our choices of hyperparameters affect the stopping point of training (based on validation loss) and vice versa.

We note that while TNG300 may be considered a medium-sized box for cosmological studies, it still should contain a few thousand voids (assuming a mean void size of $\sim 10$Mpc; \cite{Platen08Voids}) and the subcube rotation done during data augmentation effectively increases the training volume by a factor of 4. One could add more consecutive rotations and also mirror operations to further increase the size of the training dataset without risking overfitting, given the lack of rotational invariance of CNNs.

% preprocessing 
Prior to training, each mass density field is scaled (``min-maxed") to the range $[0,1]$. The min-max operation is applied to the final snapshot. In the case of processing different snapshots one could either compute a global min-max for all the snapshots used during training, or one could apply the min-max range computed from one snapshot to the rest. This scaling allows our models' training to converge more quickly and without any extreme gradients that could lead to the model failing to learn. If the input features are not scaled, some gradients calculated in the course of backpropagation can ``explode," becoming excessively large and hindering further training. Similarly, gradients can also ``vanish," which results in slow training and can often cause the model to get stuck in a suboptimal local minimum of the loss landscape. Deep neural networks such as the U-Nets we employ are especially susceptible to these effects, because gradients are multiplied across many layers during the course of backpropagation \citep{Goodfellow-et-al-2016}. 
We also examined using the logarithm of mass density, and standardization, a normalization technique common in computer vision, where the training data is normalized such that its mean is zero and standard deviation is one. We found minimal differences in predictive performance between min-max normalization and standardization and elected to use min-max.

\begin{figure}
    \includegraphics[width=0.45\textwidth]{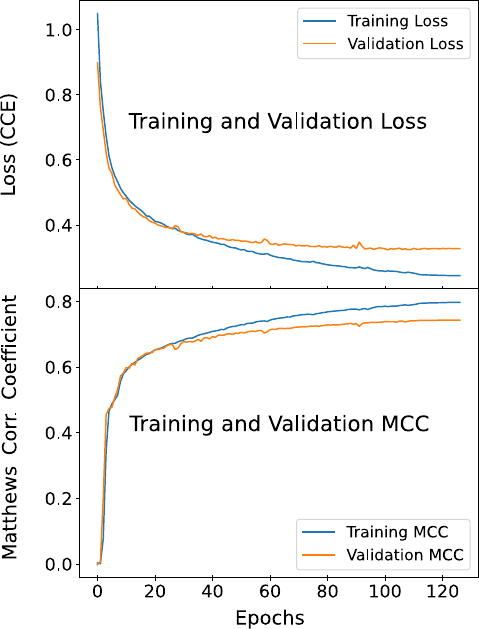}
    \caption{Metrics during the training of a DeepVoid base model. \emph{Upper:} SCCE loss vs. training epochs. \emph{Lower:} Matthews correlation coefficient for training and testing datasets during training. Note that while around epoch 40 the training and validation scores begin to diverge we further monitor the validation loss to stop training when it does not improve for 25 epochs. 
    }
    \label{fig:epochs_vs_loss}
\end{figure}

\subsection{Training}\label{subsec:training}
Each DeepVoid network was set to train for a maximum of 300 epochs, halting if the validation loss of the model had not improved in 25 epochs, and we saved the weights that resulted in the lowest validation loss, as seen in Figure~\ref{fig:epochs_vs_loss}. Early stopping is one of the chief ways that we can avoid overfitting when training models on a relatively small dataset. The average number of epochs is 85 across all trainings, with the maximum being 250. To avoid overshooting potentially better local minima in the loss landscape, we set the learning rate to quarter if the validation loss has not improved in 15 epochs. This adjustment of learning rate can also help the model fine-tune its weights more delicately in the final stage of training when the validation loss has leveled off, leading to better predictive performance and generalization.

\subsection{Curriculum learning}\label{subsec:curriculum}
Curriculum learning (CL) is a training strategy used in machine learning that involves beginning with `easier' tasks and progressively increasing the tasks' difficulty \citep{Bengio2009, soviany2021}. The idea is to have the model establish a strong foundation before attempting to classify more difficult examples. In the method of transfer learning, training of the model for the harder task begins in a location in the loss landscape that it inherits from the model trained on the easier task, then further training fine-tunes the model for the harder task. The goal of DeepVoid is to train a model that classifies regions of space based on the spatial distribution of galaxies, which act as tracers of the matter distribution. 
In this case, a model might first be trained on a relatively dense dataset (the easy task), then the weights from that model are used as the initial conditions for training a model that performs well on a sparse data set (the hard task). 

We initially train `base' models on a mass density field computed by considering all DM particles in the simulation. For TNG300-3-Dark, this results in an average interparticle separation of $\lambda = 0.33$ \hMpc. To obtain sparser samples that more closely resemble the actual distribution of galaxies that we observe at medium to high redshifts, we subsample each simulation's halos/subhalos. To do so, we take the most massive $N_{\rm subhalos}$ such that the intertracer separation is the desired value. Next, we apply the CIC interpolation scheme to the subhalos to obtain the subhalo matter density contrast as described in Section~\ref{sec:ttensor}. The multiclass truth table remains the same, as it was computed from the gravitational potential of the full density field. This allows the model to train and develop weights that can take a sparse density field as input and predict structural classifications that match the labels computed from the full density field. We created density volumes with average intertracer separations of $\lambda=[1, 3, 5, 7, 10]$ \hMpc, with most of our tests evaluating performance on $\lambda=10$ \hMpc.

\subsection{Prediction}\label{subsec:prediction}
DeepVoid computes class scores for void, wall, filament, and halo classes that sum to unity in each voxel and assigns the class with the highest score as the prediction for each voxel (see Section \ref{subsec:unet}). 
Due to GPU memory constraints, we cannot load and perform a prediction on an entire TNG mass density cube. Instead, we split the volume into overlapping subcubes of size $128^3$ that overlap by half, similar to the training process. DeepVoid then predicts in batches of eight subcubes. 
To reconstruct the full predicted volume, we reassemble the overlapping subcubes by selecting only the central region of each predicted subcube. 
This recombination process produces a seamless, nonoverlapping reconstruction of the predicted semantic segmentation with consistent predictions across subcube boundaries.\\

DeepVoid offers a significant speedup compared to the FFT-based calculation of the tidal tensor segmentation. Prediction of a $512^3$ cube takes 15 minutes on a V100 Tesla card versus 88 minutes on a Xenon server with 24 CPUs. This represents an improvement of almost 600\% which makes DeepVoid a more efficient alternative to traditional analysis. However, we note that the ultimate purpose of DeepVoid is not to speed up estimation of the tidal tensor segmentation but rather to enable the use of the predicted tidal tensor for deep galaxy redshift surveys with artifacts in which we do not have sufficient sampling density to use standard techniques.\\

% 5 analsysis of void detections using CNN
\begin{deluxetable*}{llllllll}[h]
\tablewidth{0pt}
\tablehead{\colhead{Depth} & \colhead{Filters} & \colhead{BatchNorm} & \colhead{Bal. Acc.} & \colhead{Loss} & \colhead{F1 (micro)} & \colhead{Void F1} & \colhead{MCC}}
\tablecaption{DeepVoid base model training results\label{table:base_models}}
\startdata
3     & 32      & True      & 0.78     & 0.31 & 0.91       & 0.96    & 0.81 \\ 
3     & 32      & False     & 0.80     & 0.31 & 0.90       & 0.95    & 0.79 \\ 
4     & 16      & True      & 0.80     & 0.35 & 0.90       & 0.95    & 0.79 \\ 
4     & 16      & False     & 0.76     & 0.35 & 0.88       & 0.94    & 0.76 \\ 
\enddata
\tablecomments{{DeepVoid} base models trained with SCCE loss on full DM density of the TNG simulation ($\lambda=0.33$ \hMpc). Depth is the number of layers of the U-Net and filters is the number of filters in the first convolutional layer. Batch normalization and the classification metrics are discussed in Appendix~\ref{sec:batchnorm} and \ref{sec:class_metrics}, respectively. All reported metrics are scored using the validation sets. The benefit of adding batch normalization is apparent.}
\end{deluxetable*}

\begin{comment}
\begin{deluxetable*}{lllllllll}
\tablewidth{0pt}
\tablehead{\colhead{Simulation} & \colhead{Depth} & \colhead{Filters} & \colhead{BatchNorm} & \colhead{Bal. Acc.} & \colhead{Loss} & \colhead{F1 (micro)} & \colhead{Void F1} & \colhead{MCC}}
\tablecaption{DeepVoid base model training results\label{table:base_models}}
\startdata
Bolshoi    & 3     & 32      & True      & 0.78     & 0.22 & 0.92       & 0.96    & 0.84 \\
Bolshoi    & 3     & 32      & False     & 0.80     & 0.22 & 0.91       & 0.96    & 0.84 \\ 
Bolshoi    & 4     & 16      & True      & 0.73     & 0.24 & 0.91       & 0.96    & 0.83 \\ 
Bolshoi    & 4     & 16      & False     & 0.68     & 0.31 & 0.88       & 0.94    & 0.76 \\ 
TNG        & 3     & 32      & True      & 0.78     & 0.31 & 0.91       & 0.96    & 0.81 \\ 
TNG        & 3     & 32      & False     & 0.80     & 0.31 & 0.90       & 0.95    & 0.79 \\ 
TNG        & 4     & 16      & True      & 0.80     & 0.35 & 0.90       & 0.95    & 0.79 \\ 
TNG        & 4     & 16      & False     & 0.76     & 0.35 & 0.88       & 0.94    & 0.76 \\ 
\enddata
\tablecomments{{DeepVoid} base models trained with SCCE loss on full DM density of Bolshoi and TNG. Depth is the number of layers of the U-net and filters is the number of filters in the first convolutional layer. Batch normalization and the classification metrics are discussed in Appendix~\ref{sec:batchnorm} and \ref{sec:class_metrics}, respectively. All reported metrics are scored using the validation sets. The benefit of adding batch normalization is apparent.}
\end{deluxetable*}
\end{comment}

\section{Void-Finding Analysis}\label{sec:anal}

To quantitatively evaluate the performance of the DeepVoid models, we employ a combination of standard multilabel classification metrics. All of these metrics are calculated from the confusion matrix (as shown below in Figure~\ref{fig:cm_comp}). The confusion matrix is a fundamental tool for evaluating classifiers that compares predicted class labels against the truth labels, so diagonal elements in the matrix represent correct classifications, while the nondiagonal elements represent the instances where the predicted and true class do not match. Definitions and a more detailed discussion of these metrics can be found in Appendix~\ref{sec:class_metrics}.

To evaluate model performance, we focus on 
the void-class F1 score and the Matthews correlation coefficient (MCC) as the primary indicators of model performance. We also consider the balanced accuracy (BA) and the micro-averaged F1 score.
The BA and F1 scores range from zero to one, while the MCC ranges from $-1$ to 1, where a zero MCC corresponds to random guessing. These metrics were selected because of their interpretability and robustness to class imbalance.

\subsection{DeepVoid Base Models}\label{subsec:basemodels}
% results for models trained on TNG on TNG
As discussed above, we train base models on the full-density TNG-300-3-Dark simulation, with interparticle separation of $\lambda = 0.33$ \hMpc. The best base TNG model achieves a void F1 score of 0.96 and an MCC of 0.81. Table~\ref{table:base_models} shows performance metrics of the best DeepVoid base models trained on the TNG DM particles and using SCCE loss functions. 
The upper row of Figure~\ref{fig:dpm-comp-scatter} compares the prediction of the best base model with the truth table from the tidal tensor classification; there is excellent visual agreement. 
In Figure~\ref{fig:dpm-comp-scatter} and throughout this paper, when we plot slices through the prediction cubes, we include voxels that are in both validation and training data so that one can see the contiguous structures. However, only validation data are used for the calculation of performance metrics.
In Section \ref{sec:segment} below, we quantify details of this visual comparison in the context of examining the effects of sparse sampling on the model performance.

We investigated training of models with greater depth (more convolutional layers) and a larger number of initial filters than the models in Table~\ref{table:base_models}. For these more complex models, we found equivalent or inferior results. 

Thus, the limiting factor on the models' performance, it seems, is not the number of trainable parameters, but rather the amount of training data available. 
From the base models' results, we find that batch normalization improves performance.

Future versions of DeepVoid will use larger simulation boxes, which we expect will improve the models' performance even further. Having a larger volume of training data will allow us to train models with greater numbers of trainable parameters, since the risk of overfitting the models will be reduced.

\subsection{Alternative Validation Test}
% 45 deg rotated tests:
As an additional test, 
we rotate the DM particles/halo distribution in TNG300 by 45$^\circ$ and recompute the tidal tensor mask. 
CNNs are by nature not rotationally invariant \citep{zeiler2013}.

We then run a prediction on the rotated mass density contrasts using the original models and compare the prediction to the actual rotated mask. TNG base models perform almost exactly as well on the rotated data as on the original validation data. The best models both earn MCCs of 0.81 and void F1 scores of 0.95 for the rotated dataset and 0.96 for the unchanged validation set.

\subsection{Toward Void Catalogs}

At this stage of the development of DeepVoid we have chosen not to produce ``void catalogs" that divide the void regions into a list of discrete voids. The tidal tensor method for segmenting large-scale structure into voids, walls, filaments, and halos does not by itself provide a ``truth table" of discrete structures for training the network. For voids, we know that the interior structure of voids is complex: thus the identification of individual voids requires one to specify a spatial scale on which to segment these inherently multiscale structures \citep{Aragon2013, Aragon24_HSPINE}. Geometric void finders such as, for example, VIDE \citep{Sutter15}, REVOLVER \citep{Nadathur14}, and VoidFinder \citep{ElAd97, Hoyle02} implicitly or explicitly choose a scale that is determined by the average intergalaxy spacing; below some scale, the observed underdensities are not statistically significant compared with Poisson fluctuations in the galaxy density, and so these methods ignore small-scale voids by default or by choice. In future work, we will examine criteria for creating discrete void catalogs from DeepVoid void regions.

% 6 training on sparse samples
%\section{Training on Sparser Samples}\label{sec:sparse} 
% figure: delta vs pred vs mask for L=0.33 (FULL)
% and subhalo pos vs pred for L=10.
\begin{figure*}
    \centering
    \includegraphics[width=\linewidth]{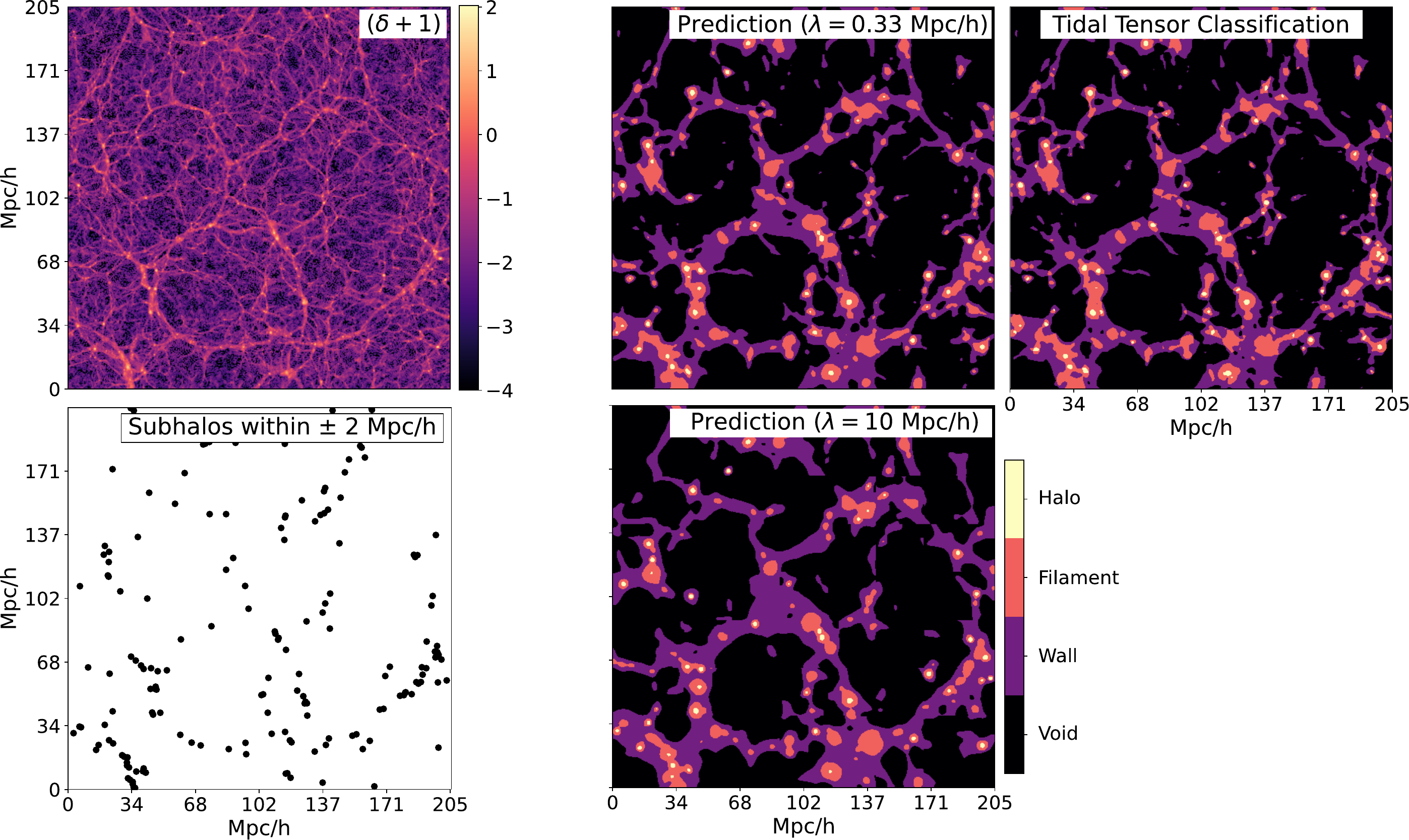}
    \caption{\emph{Upper row:} TNG mass density contrast and prediction for the best base TNG model predicting on the full TNG particle distribution with $\lambda=0.33$ \hMpc. Upper right panel is the tidal tensor truth table, which is the same for all models. \emph{Lower row:} TNG subhalo positions and prediction for the best CL model.
    Subhalos are
    first ordered by mass, then abundance matched such that the average intertracer separation is $\lambda=10$ \hMpc. Subhalos that are within 2 \hMpc of either side of the slice taken through the simulation are shown. The base model was trained on the TNG full density with a U-Net with three layers of depth and 32 initial filters. The model was then curriculum-trained on subhalos with $\lambda=3$ \hMpc average spacing 
    and then curriculum-trained again out to $\lambda=10$ \hMpc. While the slices through the prediction cubes necessarily include voxels in both the training and validation data, only the validation data are used to compute performance metrics reported in this paper.}\label{fig:dpm-comp-scatter}
\end{figure*}

\section{Training on Sparser Samples}\label{sec:sparse} 

The DeepVoid base model performs quite well, indicating that training has produced sets of
convolutional filters that identify the spatial pattern of structures that result from gravitational physics predicted by the tidal tensor.
In that first step of the DeepVoid approach, the performance of the base model indicates that the DeepVoid model is an accurate ``detector” of the predicted shapes of voids and
other structures in the full matter density field, for which DeepVoid acts as an emulator of the tidal tensor segmentation.
For application to observational data, the next step is to extend the model to detect the pattern of tidal
tensor structures in the distribution of galaxies, which are sparse, biased tracers of the matter density.

Thus, to explore building a version of DeepVoid that can be applied to density fields from deep galaxy surveys,
we must train and predict on density fields with larger intertracer spacings than the full DM particle fields of TNG. The subhalo catalog of TNG provides an adequate test case for DeepVoid, as described in Section~\ref{sec:simulations}.

\subsection{Directly Training with Sparse Samples}
% running base models on high sparsity delta:
The most naive way to predict the morphology of a volume with a large average intertracer separation would be to simply run a prediction using a base model. As expected, the base models, which are trained on density fields with intertracer spacings of much less than one, perform very poorly when predicting on the $\lambda = 10$ \hMpc subhalo mass density. The best TNG base model reaches a BA of 0.27 and an MCC of 0.07. Recalling the definition of the MCC, that score is only slightly better than random guessing.

% running models directly on high intertracer seps:
Next, we train a new model directly on the volumes with intertracer separations similar to those of galaxy redshift surveys of interest. This vastly outperforms the base models. 
Training on the subhalo density field with $\lambda = 10$ \hMpc results in scores of void F1=0.89 and MCC=0.56. As discussed in the next section, we can further improve the models' predictive performance using a multistep training process.

% figure: void-wall misclassifieds for two intertracer spacings: 
\begin{figure*}[t]
    \centering
    \includegraphics[width=\linewidth]{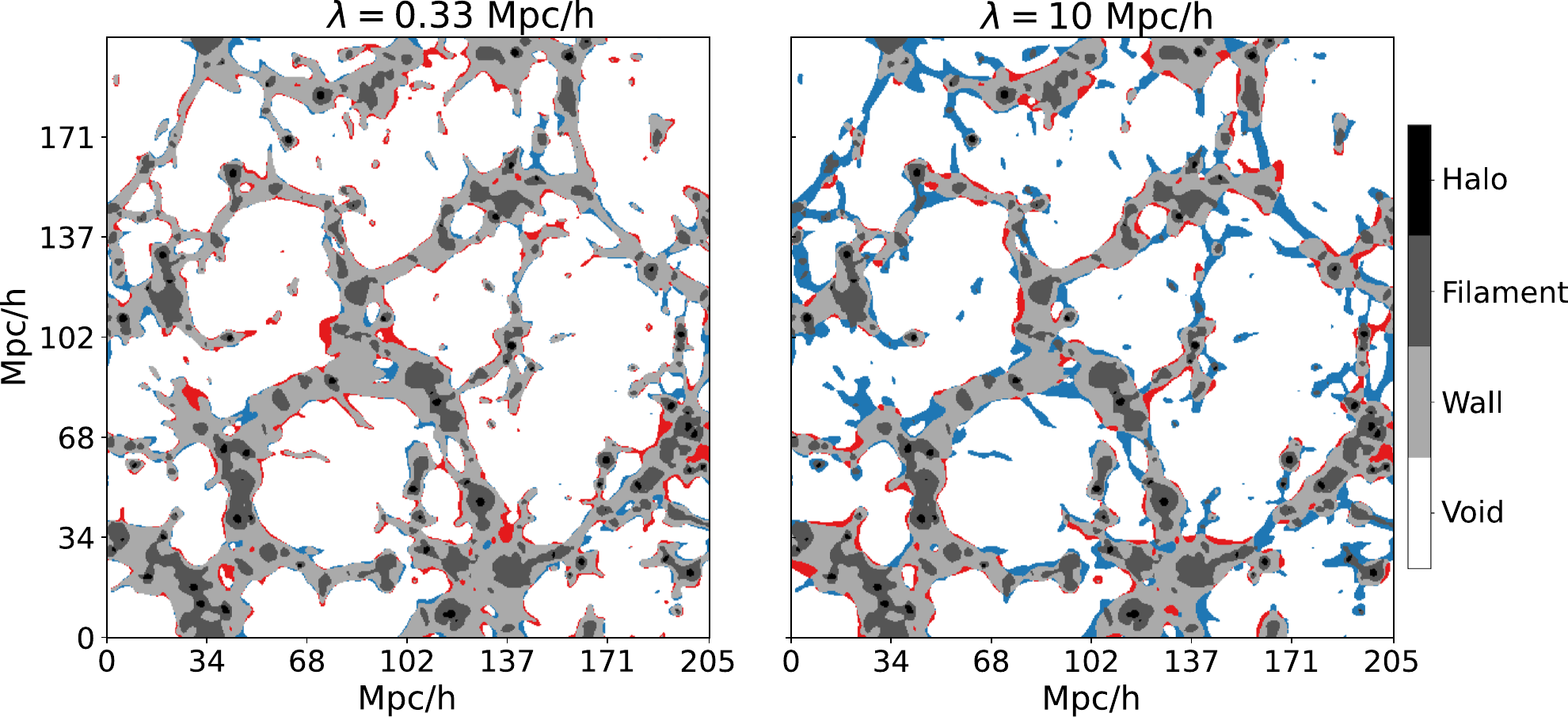}
    \caption{Slices through the TNG300-3-Dark volume. The tidal tensor mask is shown in gray-scale on both panels. Red voxels represent true void voxels misclassified as wall, while blue voxels show true wall voxels misclassified as void. \emph{Left:} base model trained on TNG dark matter particles with a 0.33 \hMpc separation. \emph{Right:} CL model trained on subhalos with a 10 \hMpc spacing. As the intertracer separation increases, the lower signal-to-noise ratio negatively impacts the quality of the segmentation, causing the model to miss small-scale features such as walls between smaller voids and tendrils of filaments extending into voids. Note that while it may appear that the percentage of red voxels does not significantly increase in the right panel, as we would expect to see from the confusion matrix in Fig. \ref{fig:cm_comp} where 4\% to 12\% are reported for the left and right panels, this appearance is due to the particular slice shown. Also, while these slices through the prediction cubes necessarily include voxels in both the training and validation data, only the validation data are used to compute performance metrics reported in this paper.}
    \label{fig:vw-misses}
\end{figure*}
% figure: confusion matrix:
\begin{figure*}
    \centering
    \includegraphics[width=\linewidth]{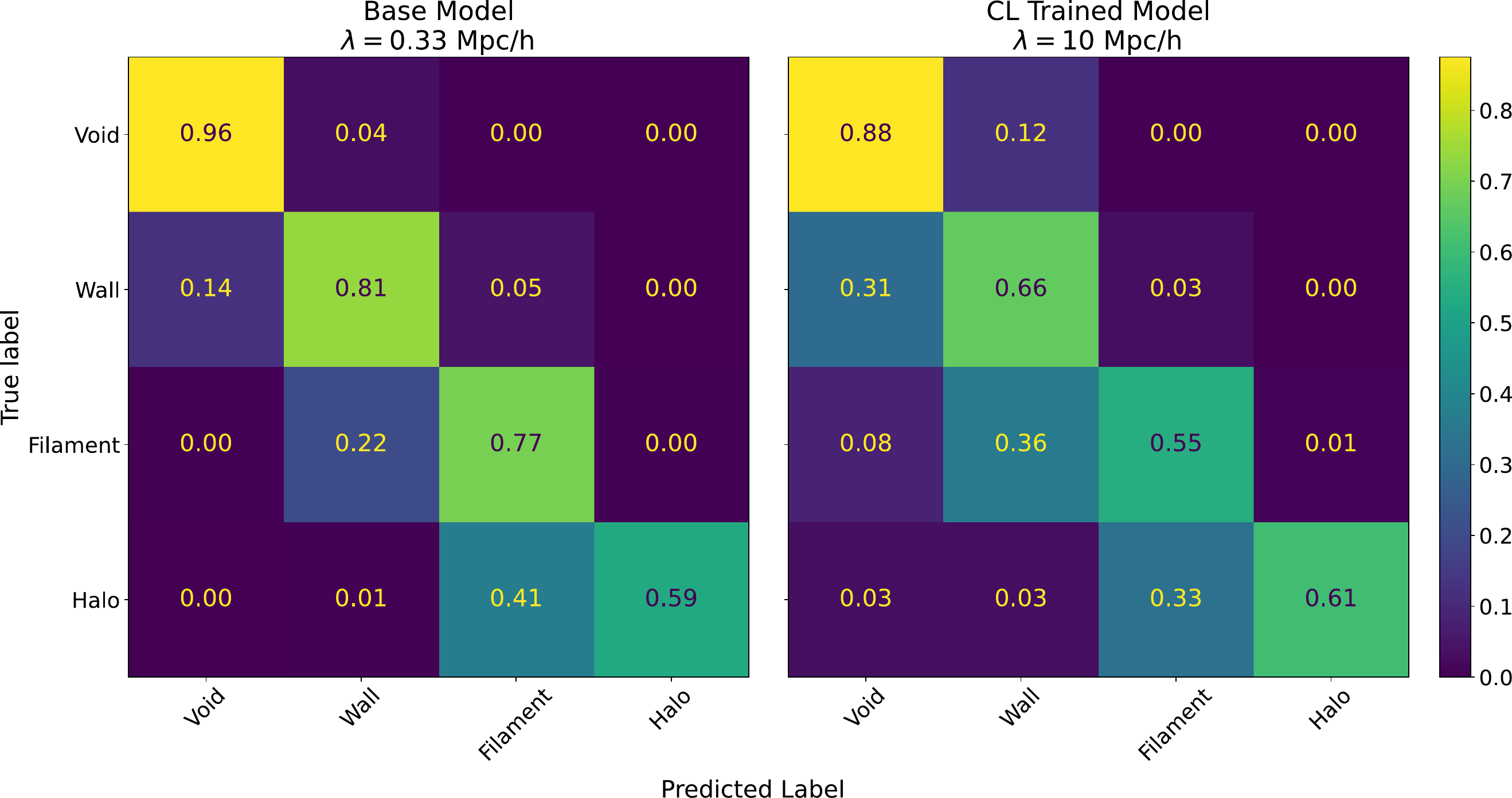}
    \caption{Comparison of two confusion matrices generated by running model predictions on the TNG validation dataset. \emph{Left:} confusion matrix from a model trained with depth of 3 and 32 initial filters on the full DM density (average interparticle spacing of 0.33 \hMpc). \emph{Right:} confusion matrix from a model with depth of 3 and 32 initial filters predicting on TNG subhalos with an average intertracer spacing of 10 \hMpc. This model was originally trained on the full TNG DM density, then curriculum-trained out to 3 \hMpc, and again to 10 \hMpc. The numbers in the matrices are normalized by the true instances, such that the rows of each matrix sum to one. The base model achieves a void F1 score of 0.96 and a MCC of 0.81 while the CL model achieves a void F1 score of 0.89 and a MCC of 0.6.}
    \label{fig:cm_comp}
\end{figure*}

\subsection{Applying Curriculum Learning}
\begin{deluxetable*}{llllllllll}
\tablewidth{0pt}
\tablehead{\colhead{Depth} & \colhead{Filters} & \colhead{BatchNorm} & \colhead{$\lambda_{\rm base}$} & \colhead{$\lambda_{\rm int}$} & \colhead{CL Type} & \colhead{Bal. Acc.} & \colhead{F1 (micro)} & \colhead{Void F1} & \colhead{MCC}}
\tablecaption{DeepVoid model results using curriculum learning\label{table:CL_models}}
\startdata
3 & 32 & True  & 0.33  & 3  & ENC\_D1\_ENC\_D2 & 0.70 & 0.81 & 0.89 & 0.60 \\ 
3 & 32 & False & 3     &    & ENC\_EO          & 0.66 & 0.81 & 0.89 & 0.60 \\ 
%3 & 32 & False & 3     &    & ENC\_EO          & 0.65 & 0.81 & 0.89 & 0.59 \\ 
3 & 32 & False & 0.33  &    & ENC\_EO          & 0.64 & 0.81 & 0.89 & 0.59 \\ 
4 & 16 & True  & 0.33  &    & ENC\_EO          & 0.67 & 0.79 & 0.87 & 0.58 \\ 
4 & 16 & False & 0.33  &    & ENC\_D1          & 0.59 & 0.80 & 0.88 & 0.57 \\ 
4 & 16 & False & 0.33  &    & ENC\_EO          & 0.64 & 0.79 & 0.88 & 0.56 \\ 
\enddata
\tablecomments{DeepVoid models that used curriculum learning to extend out to an intertracer separation of 10 \hMpc. As always, all listed scores were generated from predicting on the validation data. The type of curricular learning employed in each training is denoted by the column CL Type: ENC\_EO meaning every other convolution on the entire encoding side is frozen, ENC\_D1 and ENC\_D2 representing all convolutions on the encoder side being frozen down to a depth of two and three, respectively. A CL type of ENC\_D1\_ENC\_D2 means that the model has been curricular trained twice, once to an intermediate intertracer spacing (here, $\lambda_{int}=$ 3 \hMpc) then again to 10 \hMpc. $\lambda_{\rm base}$ represents the interparticle or intertracer spacing of the density field that the base model was trained on, while $\lambda_{\rm int}$ is the intermediate intertracer spacing, if there is one for each respective curricular learning.}
\end{deluxetable*}
As outlined in Section~\ref{subsec:curriculum}, curriculum learning aims to successively train a model on increasingly difficult examples. Typically, when training a network, the weights of the model are initialized randomly and then refined iteratively through the process of stochastic gradient descent. However, this is not the case for curriculum learning, as some of the initial weights of the model have been preserved from previously trained models that were trained on a simpler example of the task. The idea is that the model inherits valuable knowledge from earlier stages of training, then further iterates to an even better region of the loss landscape. In practice, this means that when we introduce a more difficult example to the model, such as a sparser mass density field, we freeze some of the weights that were learned by training on the `easier' (in this case, denser) example.

We experiment with different schemes for freezing the weights. 
Commonly, the weights of the encoding side of the U-net are frozen to allow for a stable feature extraction process, while the decoding side is allowed to focus on interpreting the already learned features and applying them to the new, more challenging data \citep{zaniar2022}. 
We perform curriculum learning using three different schemes: 
freezing weights in the entire encoding side (ENC),
freezing weights in only one of the two convolutional blocks on the encoding side (ENC\_EO), or freezing weights only down to some layer of depth on the encoding side (ENC\_D1/D2).
For each CL scheme, we begin with weights from the best base model,
freeze the specified set of weights, and train the unfrozen weights using training data with an intertracer separation that is larger than that on which the base model was trained.
No models with the ENC scheme outperformed the ENC\_EO models, so they are not discussed further in this analysis.

%- Add table of best CL models, Mention x2 CL procedure and results. 
Table~\ref{table:CL_models} lists performance metrics for models trained using CL. The first entry of this table characterizes a model that uses CL twice: a first step that uses CL to progress from the base model to training on an intermediate sampling of $\lambda_{\rm int}$, with weights in the first two layers of encoding side of the network frozen; and a second step of CL that trains on the final desired sampling, $\lambda_{\rm pred}$, with weights in all three layers of the encoding side frozen.

For every model in this table, $\lambda_{\rm pred}=10$ \hMpc. 
The best CL models achieve a void F1 score of 0.89 and an MCC of 0.60.
The two-step model slightly outperforms one-step CL models in BA.
The lower row of Figure~\ref{fig:dpm-comp-scatter} shows the result of training a model using this two-step CL scheme out to intertracer separation of 10 \hMpc. As with the base model, there is excellent visual agreement between the predicted class labels and the ground truth derived from the tidal tensor. In Section \ref{sec:segment} we quantify details of this comparison.

Note that Tables~\ref{table:base_models} and~\ref{table:CL_models} present only the most successful training schemes. We experimented with numerous other combinations of hyperparameters and loss functions, not shown here. These results represent a successful proof-of-concept test of classifying LSS from a sparse distribution, but are by no means exhaustive.

\begin{table}[h!] 
\centering 
\begin{tabular}{lcccc} 
\hline \textbf{Model} & \textbf{BA} & \textbf{Micro F1} & \textbf{Void F1} & \textbf{MCC} \\ \hline 
Base Binary & 0.94 & 0.95 & 0.96 & 0.88 \\ 
Base Multiclass & 0.78 & 0.91 & 0.96 & 0.81 \\ 
CL Binary & 0.81 & 0.85 & 0.89 & 0.66 \\ 
CL Multiclass & 0.70 & 0.81 & 0.89 & 0.60 \\ \hline \end{tabular} 
\caption{Comparison of classification scores: Balanced Accuracy (BA), Micro-averaged F1, Void class F1 and Matthews correlation coefficient (MCC) for Base and Curriculum Learning (CL) UNet models. CL scores are for $\lambda=10$ Mpc$/h$.} 
\label{tab:base_cl_comparison}
\end{table}

From these results, we conclude that performing CL improves the quality of the segmentation for sparser samples. In the case of voids (see Void F1 score) both binary and multiclass classifications perform identically, even at high subsampling (see Table \ref{tab:base_cl_comparison} for a comparison between scores for base and $\lambda=10$ models with and without CL). MCC and Micro F1 scores are slightly lower for multiclass models because small misclassifications within the nonvoid classes (e.g., filament vs. wall) reduce those metrics.
Using CL techniques, we improve training on subhalo distributions with intertracer spacings as large as $\lambda_{\rm pred}=10$ \hMpc.
We find that the two-step CL scheme (ENC\_D1\_ENC\_D2) yields slightly better results than a single-step CL scheme. Further exploration of model hyperparameters might yield further improvements.

\subsection{Segmentation Performance in Sparse Samples}\label{sec:segment}
Visual inspection of Fig.~\ref{fig:dpm-comp-scatter} suggests that the base and CL models accurately predict the tidal tensor classification over a wide range of tracer density, which varies by a factor of $(10/0.33)^3\approx 3\times 10^4$ in the validation samples we consider. However, a closer examination shows that the classification degrades as the tracers of the density field become sparser. This is not unexpected; it is logical that in a sparse survey, it will be more difficult to distinguish void borders and walls than in a densely sampled survey.

An increase in the number of misclassified voxels with a reduction in tracer density is highlighted in the slice plots of Figure~\ref{fig:vw-misses}. Voxel misclassification is quantified in the confusion matrix in Figure~\ref{fig:cm_comp}, where the left panel describes results for the base model trained on the full DM distribution in TNG300 and the right panel describes results for CL training on very sparse subhalos. The off-diagonal entries of the confusion matrix indicate that the proportion of wall voxels misclassified as voids rises from 14\% to 31\% when comparing the best base and CL models. At large intertracer separations, the model predictions on the edges of voids degrade, as seen in the red pixels in Figure~\ref{fig:vw-misses}. Furthermore, very small regions of walls are misclassified as voids, as shown in the blue regions of Figure~\ref{fig:vw-misses}.
Note that the exact percentages of misclassification in the particular slice shown in Figure~\ref{fig:vw-misses} do not match the statistics over the entire simulation volume, simply due to variation in the small number of such structures in a single thin slice.

\begin{figure}
    \centering
    \includegraphics[width=\linewidth]{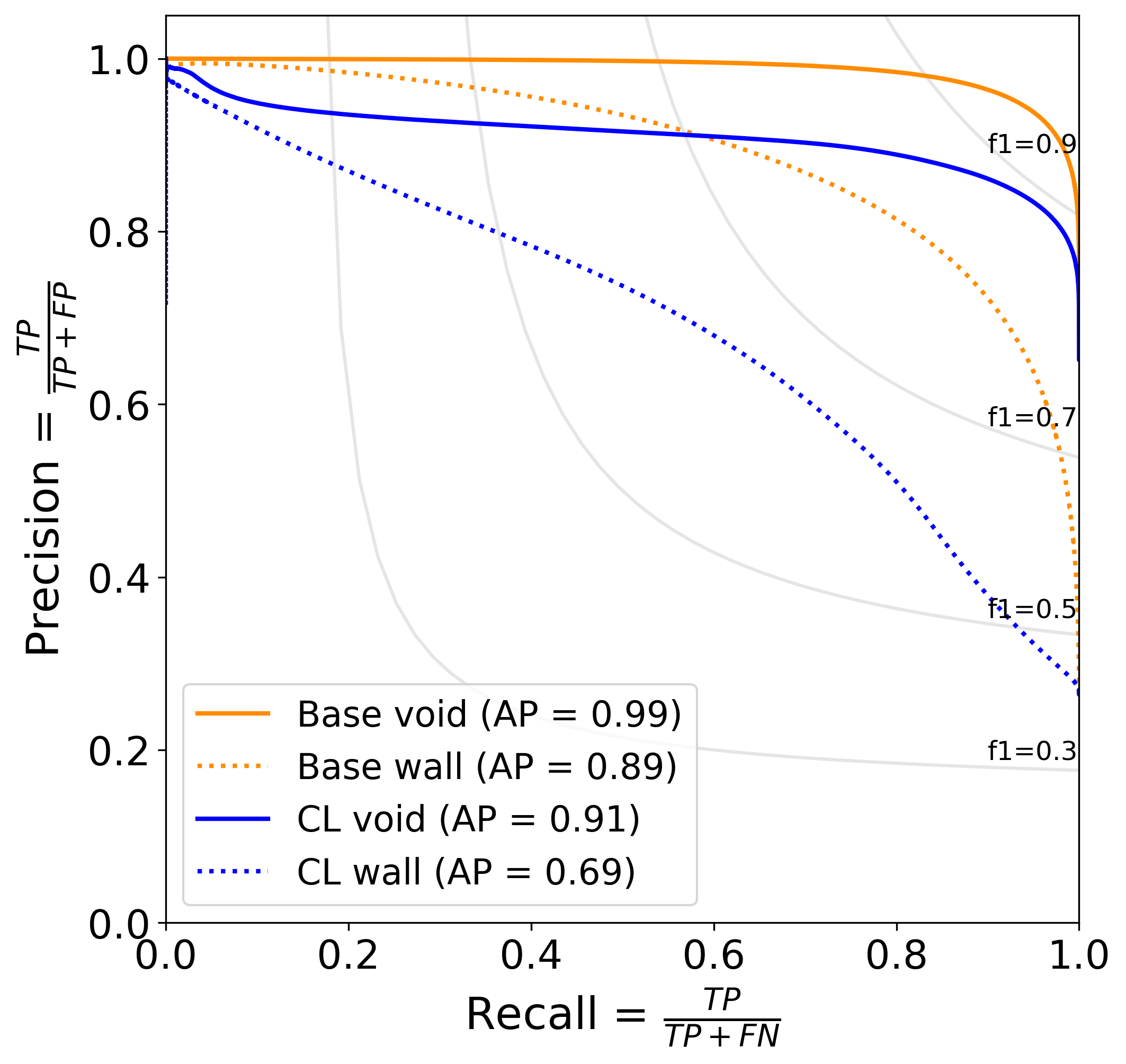}
    \caption{Precision-Recall (PR) curves of void and wall classification for base and CL models with intertracer spacings $\lambda=0.33$ \hMpc and $\lambda=10$ \hMpc, respectively. Grey lines represent lines of constant F1 score (harmonic mean of precision and recall) for reference. The legend includes the average precision (AP) of each case. The models' performance on the void class across the large difference of intertracer separation is remarkably consistent. However, performance on wall classification degrades significantly, as seen in this figure and in the confusion matrix (Figure~\ref{fig:cm_comp}).}\label{fig:PR_CURVE}
\end{figure}

The precision-recall (PR) curves in Figure~\ref{fig:PR_CURVE} further quantify the performance of the models in detecting voids and walls, and allow us to visualize the trade-off between high classification purity (precision) and classification efficiency (recall). 

PR curves take advantage of the full distribution of class-specific activation scores. 
For example, to compute the void PR curve, 
we vary the void-class score from 0 to 1, using
the same score across the entire simulation cube to separate void voxels from nonvoid voxels.
This separation of void voxels from nonvoid voxels is different from the multiclass segmentation above where each voxel was assigned the class with the highest score for that voxel alone. 

For each void-class score used to separate void voxels from nonvoid voxels, we compute the precision -- defined as the proportion of correctly identified void voxels out of all predicted void voxels -- and the recall -- defined as the fraction of true void voxels identified. Varying the class threshold along the PR curve, one obtains perfect precision at an extremely high threshold (false positives approach zero), and perfect recall at an extremely low threshold (false negatives approach zero).
F1 scores (grey curves in Figure \ref{fig:PR_CURVE}) measure the balance between precision and recall.
Appendix~\ref{sec:class_metrics} provides further details of the precision and recall metrics.

The CL model accurately classifies void voxels even with a large increase in intertracer spacing. This is shown by
the relatively stable void PR curve, which  indicates that the model remains effective at void identification, with the CL-trained model achieving an average precision $\text{AP} =0.91$ and $\text{F1} =0.89$ (harmonic mean of precision and recall) for voids in sparse data ($\lambda=10$ \hMpc).
In contrast, the average precision for walls drops significantly compared to the base model. The corresponding confusion matrices suggest that this drop in wall precision reflects an increase in the fraction of voids predicted as wall from 4\% to 12\%, and a rise in the fraction of filaments predicted as wall from 22\% to 36\%. Thus, the PR curves and confusion matrices quantify the degradation of predictions along void boundaries observed in Figure~\ref{fig:vw-misses}.

% PR Curve Figure

% PR Curve Figure
\begin{figure}
    \centering
    \includegraphics[width=\linewidth]{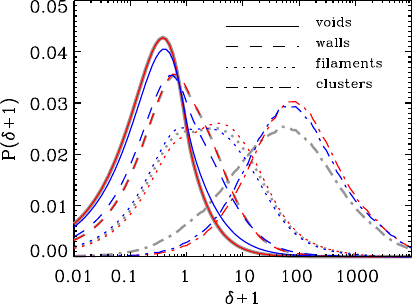}
    \caption{Density distribution of voxels inside voids, walls, filaments, and clusters. The training mask is shown in grey while the predictions for the $\lambda=3$\hMpc and $\lambda=10$\hMpc samples are indicated by red and blue lines respectively.}\label{fig:density_dist}
\end{figure}

Figure \ref{fig:density_dist} shows a comparison between the density distribution inside the true and predicted masks. It serves as a first test of the physical validity of our results. The density distributions follow the expected, behavior increasing from voids to walls, filaments and clusters. While each cosmic environment has a distinct peak, there is a large overlap between the distributions that highlights the difficulty in separating environments by density alone. The stable performance of DeepVoid, shown in the similarity of the three samples, even in the sparsest sample, is encouraging. However, as shown in Figs. \ref{fig:vw-misses} and \ref{fig:cm_comp}, at low sampling density ($\lambda=10$ \hMpc), some void voxels near the peak and at slightly lower density are misclassified as relatively low-density wall voxels, while some wall voxels at density above the wall peak are misclassified as relatively high-density void voxels. This is not unexpected, as the low sampling decimates the structures DeepVoid is seeking to identify.

% 7 discussion

\section{Conclusions}\label{sec:disc}
% summary
In this work, we present DeepVoid, a novel multiclass method of semantically segmenting the morphology of structure in cosmological simulations. We define four classes of large-scale structures based on the tidal tensor computed from the gravitational potential: voids, walls, filaments, and halos \citep{Hahn2007}. We then train deep convolutional neural networks to reproduce this truth table, given a density contrast field as input. We train U-Net models on the TNG300-3-Dark simulation. When trained on the full DM particle density field, with interparticle separation $\lambda=0.33$ \hMpc, the best model achieves an MCC of 0.81 and a void F1 score of 0.96 on validation data.

We use curriculum learning to extend the models' performance to sparse samples of subhalos in the simulations. For the lowest density of tracers that we trained on, with intertracer spacing of $\lambda=10$ \hMpc, the best CL-trained model scores a MCC of 0.60 and a void F1 of 0.89. The performance of DeepVoid models on such sparse samples is promising for future void detection in high-redshift galaxy surveys.

While DeepVoid employs a physical definition of structure, some parameter choices still influence the segmentation results. The eigenvalue threshold for the tidal tensor has the most significant impact on structural classification. Setting it to zero restricts void classification to only the most underdense centers of voids, while larger thresholds expand void regions substantially, attenuating the connections between walls and filaments. We select a physically motivated threshold that also matches previously observed volume fractions of voids.

This prototype employs a convolutional neural network in a U-Net architecture, for which we explored variations of hyperparameters. Further increasing model depth and filter count did not result in increased performance for training using training data from TNG300. 
Further refinements in training on larger simulation volumes beyond TNG300 represent natural extensions to this work.
Future work could also explore novel architectural elements such as residual networks, self-attention mechanisms, and vision transformers \citep{oktay2018,Liu2023,dosovityskiy2020,chen2024}, all of which have shown promise in analogous computer vision tasks. 

It is important to note that while DeepVoid was developed to detect cosmic voids in sparse datasets, it is also fully capable of identifying regions in walls, filaments and clusters (see Fig. \ref{fig:dpm-comp-scatter}). The confusion matrix (Fig. \ref{fig:cm_comp}) shows the identification of regions in walls, filaments, and clusters even in sparse galaxy surveys. DeepVoid can enable environmental studies of galaxies in the cosmic web up to redshifts where other methods fail due to poor sampling of cosmic regions. The U-Net at the core of DeepVoid learns patterns in the density field and translates them to labels even for highly sparse samples. This is aided by the use of curriculum learning, in which the density field is progressively degraded, and the patterns are learned based on previous training.

In the follow-up work to this paper, we will develop versions of DeepVoid that are trained on mock galaxy redshift surveys from simulations.
The identification of voids from biased tracers requires careful consideration, as the observed galaxy distribution can be affected by the background cosmology and (in the case of simulations used for training) the galaxy formation physics of the simulation or semianalytic model used to construct galaxy catalogs.
These applications of DeepVoid will take as input survey-format data and deal with all the accompanying complications, such as redshift-space distortions and handling survey mask boundaries. The redshift-space distortions can also be leveraged to improve training and, therefore, our predictions, since there is information inherent in the orientation of structures relative to the line of sight. Future iterations of DeepVoid will also incorporate galaxy properties that are known to vary with environment, such as color and luminosity, to examine how environmental dependence may improve void detection.
Once DeepVoid models are trained on simulated survey data, the next step will be to apply them to galaxy survey data from SDSS \citep{SDSS7}, SDSS BOSS \citep{Dawson2013}, and DESI \citep{DESI2016a.Science} to produce void catalogs suitable for cosmological tests.

% 8 discussion and conclusions and future work
\bigskip
\centerline{\bf Acknowledgements}
\smallskip

We thank the anonymous referee for the suggestions, which
substantially improved this paper.
Primary computing for this paper was done on Picotte, which is supported by the Drexel University Research Computing Facility.
Additional computing for this work was supported in part by the National Research Platform Nautilus cluster, which is supported in part by National Science Foundation (NSF) awards CNS-1730158, ACI-1540112, ACI-1541349, OAC-1826967, OAC-2112167, CNS-2100237, CNS-2120019, the University of California Office of the President, and the University of California San Diego’s California Institute for Telecommunications and Information Technology/Qualcomm  Institute, a Consejo Nacional de Humanidades Ciencia y Tectnología (CONAHCyT) Ciencia de Frontera grant CF-2023-1-1971 and Universidad Nacional Autónoma de México (UNAM) Programa de Apoyo a Proyectos de Investigación e Innovación Tecnológica (PAPIIT) grant IN115224. Thanks to CENIC for the 100Gbps networks.

This project was made possible through the support of grant 62177 from the John Templeton Foundation. The opinions expressed in this publication are those of the authors and do not necessarily reflect the views of the John Templeton Foundation.

\bigskip
\centerline{\bf Data availability}
\smallskip

The code used in this paper is publicly available at the following link: \url{https://github.com/skumagai3/deepvoid_misc}. The data used to train and test DeepVoid can be provided upon reasonable request of the authors.

%\software{Tensorflow}
\textit{Software:} Tensorflow \citep{tensorflow}, Keras \citep{keras}.

% 9 appendix: batch norming, metrics
\appendix
\label{sec:Appendix}

% Tidal tensor detailed derivation
\section{Tidal Tensor Derivation}\label{sec:tt_derive}
To apply our eigenvalue-based segmentation scheme to DM particles in a simulated box, we first CIC the particle distribution to obtain the matter density contrast as described in Section~\ref{sec:ttensor}. We can then relate this to the potential by use of the 3D Poisson equation:
\begin{equation}
    \nabla^2 \Phi(\Vec{r}) = 4 \pi G \rho(\Vec{r}),
\end{equation}
where $G$ is the gravitational constant and $\rho(\Vec{r})$ is the matter density in real space. We can use a Green's function to make this equation more manageable, as it is defined as the solution of 
\begin{equation}
    \nabla^2 \mathcal{G}(\Vec{r}) = \delta_D(\Vec{r}),
\end{equation}
or equivalently in Cartesian coordinates:
\begin{equation}
    \nabla^2 \mathcal{G}(x,y,z) = \delta_D(x)\delta_D(y)\delta_D(z),
\end{equation}
where $\delta_D$ in this case represents the Dirac delta function. It follows that the Green's function, the solution of the above equation, is written as
\begin{equation}
    \mathcal{G}(\Vec{r}) = -\frac{1}{4 \pi |\Vec{r}|}.
\end{equation}
We can then rewrite the Poisson equation as
\begin{equation}
    \Phi(\Vec{r}) = 4 \pi G (\mathcal{G} * \rho)(\Vec{r}),
\end{equation}
where $*$ denotes the convolution product between the Green's function and the matter density. Because convolutions in configuration space are multiplications in Fourier space, we can apply a forward Fourier transform to obtain
\begin{equation}
    \hat{\Phi}(\Vec{k}) = 4 \pi G\ \hat{\mathcal{G}}(\Vec{k}) \hat{\rho}(\Vec{k}),
\end{equation}
where $\Vec{k}$ is the wavevector and hats denote values in Fourier space. We can easily calculate Green's function in Fourier space as
\begin{equation}
    \hat{\mathcal{G}}(\Vec{k}) = -\frac{1}{|\Vec{k}|^2}.
\end{equation}
Then, for $\hat{\Phi}(\Vec{k})$:
\begin{equation}
    \hat{\Phi}(\Vec{k}) = - 4 \pi G \frac{\hat{\rho}(\Vec{k})}{|\Vec{k}|^2}.
\end{equation}
Finally, we perform an inverse Fourier transform to obtain the gravitational potential in real space, $\Phi(\Vec{r})$. A Gaussian kernel then smoothes this potential. We then calculate the Hessian matrix:
\begin{equation}
    \left( \frac{\partial^2 \Phi}{\partial x_i \partial x_j} \right) ,
\end{equation}
and compute the eigenvalues for each voxel. Finally, the morphological segmentation is generated by choosing a threshold for the eigenvalues, \lth, and applying it to all voxels in the periodic box.

\section{Classification Performance Metrics}\label{sec:class_metrics}
% moved here from analysis
Most classification metrics are some combination of the elements of the confusion matrix, which we define in Section~\ref{sec:anal}. To see examples of confusion matrices, see Figure~\ref{fig:cm_comp}.

% describe why accuracy is not a good metric for us
Accuracy is one of the most commonly used and easily understood metrics for evaluating any predictive process. In particular, accuracy is defined as the ratio of correctly classified instances to the total number of instances:
\begin{equation}
    \text{accuracy} = \frac{\mathrm{TP} + \mathrm{TN}}{\mathrm{TP} + \mathrm{TN} + \mathrm{FP} + \mathrm{FN}}\ ,
\end{equation}
where TP represents true positives, TN true negatives, FP false positives, and FN false negatives. In our case, we calculate the global accuracy of our classifier by macroaveraging, i.e. calculating the accuracy for each class and averaging the four scores. 
However, it is misleading to use accuracy as a metric for imbalanced datasets. In our case, where we have many more void voxels than filament or halo voxels (61\% vs. 5\% and 0.5\% of the volume, respectively), the model could simply predict void for every voxel and reach what looks to be a reasonable level of accuracy, since false positives are not penalized. 

% define balanced accuracy 
The balanced accuracy quantity is specifically designed to address these issues with imbalanced datasets. The balanced accuracy is the mean of the true positive rate (sensitivity or recall) and the true negative rate (specificity) across all classes. The recall represents the probability of a positive instance being correctly classified as positive, while specificity is the probability of a negative instance being correctly classified as negative. The balanced accuracy therefore takes the form:
\begin{equation}
    \text{balanced acc.} = \frac{1}{2} \left( \frac{\mathrm{TP}}{\mathrm{TP}+\mathrm{FN}} + \frac{\mathrm{TN}}{\mathrm{TN}+\mathrm{FP}} \right) .
\end{equation}

% define precision, recall
Another relevant classification metric for our purposes is the precision, which represents the ratio of true positives to all positive predicted instances by the classifier:
\begin{equation}
    \text{precision} = \frac{\mathrm{TP}}{\mathrm{TP}+\mathrm{FP}}\ .
\end{equation}
The recall, which was referenced above, is the ratio of true positives to all positive instances:
\begin{equation}
    \text{recall} = \frac{\mathrm{TP}}{\mathrm{TP} + \mathrm{FN}}\ .
\end{equation}
% Dice score F1 for imbalanced classes
An important metric for evaluating our models' performance is the Sørenson-Dice coefficient or Dice score, which is also known in machine learning circles as the F1 score \citep{dice1945, Srensen1948}. The Dice score is the harmonic mean of precision and recall. For a binary classifier, it can be expressed as 
\begin{equation}
        \begin{aligned}
            F_1 & = 2 \times \frac{\text{precision} \times \text{recall}}{\text{precision} + \text{recall}}\ , \\
            F_1 & = \frac{2\times \mathrm{TP}}{2 \times \mathrm{TP} + \mathrm{FP} + \mathrm{FN}}\ .
        \end{aligned}
\end{equation}
The F1 score is a useful metric for evaluating classifiers since it combines precision and recall into a single value, providing a balanced assessment of a model's performance. Furthermore, it is easily interpretable as it ranges from zero to one, and provides an easy way to compare models. Since we are generally more concerned with finding voids in this work, we pay special attention to the void-class F1 score.

% macro vs. micro averaging of F1 scores.
However, as this is a multiclass problem, we also want to evaluate our models' prediction quality across all classes. There are several ways to average across classes: micro, macro, and weighted averaging. Microaveraging calculates the F1 score globally by considering the total number of TP, FP, and FN across all classes. This approach is useful for dealing with imbalanced datasets, since it ensures that each instance contributes equally to the final F1 score. In contrast, macroaveraging simply consists of computing the F1 score for each class and averaging those scores, giving equal weight to each class. 
Weighted averaging is computed similarly to the macroaverage, but weights each score by its ``support," or the number of true instances for each class. However, this may result in an F1 score that does not lie between the precision value and the recall value, obscuring the original meaning of the F1 metric. 

% (F1 drawbacks: doesn't consider true negatives)
% Matthew's correlation coefficient
One drawback of the F1 score is that TN are never considered in the calculation of the metric, and, therefore, it could be misleading for datasets with class imbalance such as ours. A metric that is often touted as a superior alternative is the Matthews correlation coefficient \citep[MCC;][]{chicco2020MCC}. The MCC was originally defined by \citet{matthews1975MCC} but was reintroduced in a machine learning context by \citet{baldi2000MCC}. The MCC provides a holistic view of the predictive performance of a classifier, even with imbalanced classes. It can be expressed by terms from the confusion matrix as:
\begin{equation}
    \text{MCC} = \frac{\mathrm{TP} \times \mathrm{TN} - \mathrm{FP} \times \mathrm{FN}}{\sqrt{(\mathrm{TP}+\mathrm{FP})(\mathrm{TP}+\mathrm{FN})(\mathrm{TN}+\mathrm{FP})(\mathrm{TN}+\mathrm{FN})}}\ .
\end{equation}
It ranges from $-1$ to $+1$, and can generally be thought of as a correlation coefficient with values at or near $-1$ meaning complete disagreement between prediction and ground truth, 0 representing random guessing, while a score of $+1$ represents a perfect prediction. To receive a good MCC, a model must perform well on the positive and negative elements for each class. 

Due to these considerations, the metrics that most accurately reflect the performance of DeepVoid models and that we use to differentiate between them are balanced accuracy, microaveraged F1, void F1, and the MCC. In particular, the MCC succinctly captures the general quality of the segmentation across all classes, and the void F1 reflects how well we segment void regions, which are the primary concern of this work.

% Batch Normalization
\section{Batch normalization}\label{sec:batchnorm}
An additional component we add to our U-Nets that improves performance is batch normalization. First proposed by \citet{2015arXiv150203167I}, batch normalization is meant to make training deep neural networks both faster and more stable by reducing the internal covariate shift. This term, while not strictly defined in the machine learning community, refers to the phenomenon where the distribution of inputs to each layer changes during training due to randomness in both the initialization of weights and inherent variability in the input data itself. Batch normalization aims to address this by normalizing the activations within each ``minibatch" of the training set. It does this by computing the mean and the variance of the feature images, then scaling and shifting the normalized values using trainable parameters. If we take a minibatch $\{ x_1, x_2, ..., x_m \}$ from a layer in the network, the minibatch mean $\mu_B$ and the variance $\sigma^2_B$ are defined as:
\begin{equation}
    \mu_B = \frac{1}{m} \sum^m_{i=1}x_i,
\end{equation}
\begin{equation}
    \sigma^2_B = \frac{1}{m} \sum^m_{i=1} \left( x_i - \mu_B \right)^2 .
\end{equation}
The normalized value $\hat{x_i}$ is then defined as
\begin{equation}
    \hat{x_i} = \frac{x_i - \mu_B}{\sqrt{\sigma^2_B + \epsilon}} ,
\end{equation}
where $\epsilon$ is to prevent division by zero. Finally, the normalized value is scaled and shifted by two learnable parameters $\gamma$ and $\beta$:
\begin{equation}
    y_i = \gamma \hat{x_i} + \beta.
\end{equation}
This normalization has been proven in a variety of applications to repress exploding gradients and to help prevent overfitting through its slight regularization effect. These effects have led to batch normalization becoming a common tool in deep learning. Typically, the batch normalization layers are placed after the convolutional layer and before the activation function in each convolutional block. This placement ensures that the activations passed to subsequent layers are normalized. However, in our U-Nets, we found that placing batch normalization layers in every block actually caused exploding gradients, harming classification quality and sometimes preventing the model from ever meaningfully learning. We experimented with other placements and found that a batch normalization layer in every other block helped training significantly.
Additionally, we tested other common network modifications such as L1 and L2 regularization and dropout layers, but neither impacted model performance positively.

\section{Alternative Loss Function}\label{sec:loss_fxns}
% Other loss functions:
While CCE is the most commonly used loss function for classification, there are several others that can be employed for multiclass classification. The focal loss, a form of CCE with two additional parameters $\alpha$ and $\gamma$, was created to address classification problems where one or more of the constituent classes is vastly overrepresented \citep{Lin2017FOCAL}. The focal loss takes the form 
\begin{equation}
    L_\mathrm{FOCAL}(y,\hat{y}) = - \sum_{i}^N \alpha_i \left( 1 - \hat{y_i} \right)^{\gamma} \log(\hat{y_i}) ,
\end{equation}
where $y$ and $\hat{y}$ are the same as in the CCE function, $\alpha_i$ is a weight for class $i$, and $\gamma$ is a focusing parameter. The main difference between CCE and the focal loss is that CCE treats all classes equally and does not explicitly address class imbalance in the training dataset. The focal loss parameter $\gamma$ down-weights ``easy" examples, i.e. predictions that the model confidently classifies correctly, and focuses more on ``hard" examples. For datasets such as ours that have large class imbalances, the idea is that this will change the shape of the loss `landscape' and ideally lead to improved performance. We tested several values of $\alpha$ and $\gamma$, starting with the default values of $\alpha=0.25$ (for all classes) and $\gamma=2$. Additionally, we set the weights of the $\alpha$ parameter to roughly reflect the balance of the classes in the datasets, so $\alpha = [0.6, 0.25, 0.15, 0.1]$. Finally, we set the $\alpha$ weights to reflect which classes were most important to classify correctly, the void and wall classes. Therefore, we also set $\alpha = [0.6, 0.5, 0.25, 0.25]$. For each of these cases, we also tested setting $\gamma=3$. While all of these models provide good segmentations, no base model with focal loss outperforms the SCCE versions.

We also examined using a combination of the CCE loss function and the Dice (F1) score \citep{galdran2022}. CCE provides a strong baseline for voxel-by-voxel classifications that typically have stable gradients and easily extend to multi-label problems. The F1 score is much more robust to class imbalance and prioritizes correct shapes and boundaries. We found that models trained using a multiclass version of the F1 score often fail to get traction on the problem and do not train. Many weightings of these two loss functions are used in the literature. 
We tested using the average of the two. We found that using this combination loss resulted in base models that either matched or very slightly outperformed the SCCE models, 
but performed slightly worse than SCCE when using CL to train on sparser data.

%%%%%%%%%%%%%%%%%%%%%%%%%%%%%%%%%%%%%%%%%%%%%%%%%%%%%%%%%%%%%%%%%%%%%%%%%%%%%%%%
% References
%-------------------------------------------------------------------------------
\bibliographystyle{aasjournal}
\bibliography{sambib,references}
%%%%%%%%%%%%%%%%%%%%%%%%%%%%%%%%%%%%%%%%%%%%%%%%%%%%%%%%%%%%%%%%%%%%%%%%%%%%%%%%
\end{document}